\documentclass[12pt]{article}

\usepackage{amsfonts}
\usepackage{amsmath}
\usepackage{epsfig}
\usepackage{latexsym}
\usepackage{graphicx}
\usepackage{subfigure}
\usepackage{amssymb}
\usepackage{cite}
\numberwithin{equation}{section}
\allowdisplaybreaks

\renewcommand{\thefootnote}{\fnsymbol{footnote}}
\def\CR{\nonumber \\}
\def\eq#1{(\ref{#1})}
\def\[#1\]{\begin{align}#1\end{align}}
\def\hM{{\hat M}}
\def\hP{{\hat P}}
\def\hH{{\hat {\cal H}}}
\def\hJ{{\hat {\cal J}}}
\def\hD{{\hat {\cal D}}}
\def\hC{{\hat {\cal C}}}

\def\hN{{\cal N}}

\def\spa#1{\phantom{\fbox{\rule[-#1cm]{0cm}{0cm}}}}

\def\be{\begin{equation}}
\def\ee{\end{equation}}
\def\bea{\begin{eqnarray}}
\def\eea{\end{eqnarray}}

\renewcommand{\thefootnote}{\fnsymbol{footnote}}

\begin{document}

\hfuzz=100pt
\title{
\begin{flushright} 
\vspace{-1cm}
\small{WITS-CTP-150, YITP-14-73}
\end{flushright} 
{\Large 
\bf Physical states in the canonical tensor model \\
from the perspective of random tensor networks
}
}
\date{}
\author{Gaurav Narain$^a$\footnote{gaunarain@nu.ac.th}, Naoki Sasakura$^b$
\footnote{sasakura@yukawa.kyoto-u.ac.jp} and Yuki Sato$^{c}$\footnote{Yuki.Sato@wits.ac.za}
  \spa{0.5} \\
$^a${\small{\it The Institute for Fundamental Study ``The Tah Poe Academia Institute", }}
\\ {\small{\it  Naresuan University, Phitsanulok 65000, Thailand     }}
\spa{.5}\\
$^{b}${\small{\it Yukawa Institute for Theoretical Physics,}}
\\ {\small{\it  Kyoto University, Kyoto 606-8502, Japan}}
\spa{.5}\\
$^{c}${\small{\it National Institute for Theoretical Physics, }}
\\ {\small{\it School of Physics and Centre for Theoretical Physics, }}
\\ {\small{\it University of the Witwartersrand, WITS 2050, South Africa}}
}
\date{}

\maketitle
\centerline{}

\begin{abstract}
Tensor models, generalization of matrix models, are studied aiming for quantum gravity in dimensions larger than two.
Among them, the canonical tensor model is formulated as a totally constrained system with first-class constraints, the algebra 
of which resembles the Dirac algebra of general relativity. 
When quantized, the physical states are defined to be vanished by the quantized constraints.
In explicit representations, the constraint equations are a set of partial differential equations 
for the physical wave-functions,
which do not seem straightforward to be solved due to their non-linear character.
In this paper, after providing some explicit solutions for $N=2,3$, 
we show that certain scale-free integration of partition functions of statistical systems on random networks
(or random tensor networks more generally) provides a series of solutions for general $N$.
Then, by generalizing this form, we also obtain various solutions for general $N$.
Moreover, we show that the solutions for the cases with a cosmological constant 
can be obtained from those with no cosmological constant for increased $N$.
This would imply the interesting possibility that a cosmological constant can always 
be absorbed into the dynamics and 
is not an input parameter in the canonical tensor model.  We also observe the possibility of symmetry enhancement in $N=3$,
and comment on an extension of Airy function related to the solutions.
\end{abstract}

\renewcommand{\thefootnote}{\arabic{footnote}}
\setcounter{footnote}{0}

\newpage
\section{Introduction}
\label{realIntroduction}
Tensor models were initially introduced as models for quantum gravity 
in $D>2$ dimensions \cite{Ambjorn:1990ge,Sasakura:1990fs,Godfrey:1990dt}, 
by generalizing the matrix models \cite{DiFrancesco:1993nw}, 
which are known to consistently describe the $D=2$ dimensional quantum gravity.
The main idea of the tensor models was that the Feynman diagrams obtained 
via perturbative treatment of the tensor models could be identified with the dual 
diagrams of simplicial manifolds for $D>2$ dimensions, the summation 
of which would evaluate in a continuum limit
the geometric sum in Euclidean quantum gravity.
However, these original tensor models were unsuccessful due 
to singular or un-oriented Feynman diagrams \cite{Sasakura:1990fs,DePietri:2000ii}, 
and also it was not clear how to take continuum limits, because of the absence of $1/N$ 
expansions\footnote{$N$ denotes the dimension of the vector space associated to the indices of 
vectors, matrices and tensors in this paper. 
Namely, an index runs through $1, 2, \cdots, N$.},
which played essential roles in taking the continuum limits of the matrix models. 
This situation has been drastically changed by the advent of 
the colored tensor models \cite{Gurau:2009tw}.  
The colored tensor models have a one-to-one correspondence 
between the Feynman diagrams and pseudo-simplicial manifolds, and
also have $1/N$ expansions, the leading terms of which are composed 
of melonic diagrams \cite{Gurau:2011xp}. 
The melonic diagrams correspond to simplicial manifolds with 
spherical topology, but are geometrically singular 
\cite{Gurau:2011xp,Bonzom:2011zz,Gurau:2013cbh}.  
Therefore it seems important to modify the leading order or 
to include higher order $1/N$ corrections to obtain continuum limits 
of physical interests as in the matrix models. A number of 
efforts are currently being made in these directions 
\cite{Raasakka:2013eda,Dartois:2013sra,Kaminski:2013maa,
Bonzom:2013lda,Gurau:2013pca,Dartois:2013he,Bonzom:2012wa,Tanasa:2011ur,Dartois:2014hga,Nguyen:2014mga}. 
The colored tensor models have also stimulated the developments of the renormalization group procedure  
of group field theories \cite{Carrozza:2014rba,Rivasseau:2013uca,Carrozza:2013mna,
Geloun:2013zka,Geloun:2013saa,Samary:2013xla,Carrozza:2013wda,
Carrozza:2012uv,Geloun:2012bz,BenGeloun:2012pu,BenGeloun:2011rc},  which 
are tensor models with group-valued indices \cite{Boulatov:1992vp,Ooguri:1992eb}
and are studied extensively in the context of the 
loop quantum gravity \cite{DePietri:1999bx,Freidel:2005qe,Oriti:2011jm}.  

The developments above of tensor models concern the Euclidean case. 
In quantum gravity with geometric fluctuations, it would be unclear
whether the time-like direction can be treated in the same way as 
the other space-like directions, which is the standard procedure with
Wick rotation in field theories on flat space-times.
In fact, in the dynamical triangulation model of quantum gravity, 
it has been shown that simplicial complexes which can be regarded as smooth space-times in global scale
dominate in Causal Dynamical Triangulation \cite{Ambjorn:2004qm}, 
while Dynamical Triangulation, the original Euclidean model, 
does not seem successful in this respect.
\footnote{When coupling many U$(1)$-fields, 
the authors in \cite{Horata:2000eg} found a promise of 
a phase transition higher than first order, 
which, however, is in conflict with the result in \cite{Ambjorn:1999ix}.}
Inspired by this fact, one of the present authors proposed 
Hamilton formalism of tensor models, dubbed the canonical tensor model 
in short \cite{Sasakura:2011sq,Sasakura:2012fb,Sasakura:2013gxg}.
\footnote{An operator formalism of group field theories has been 
developed in \cite{Oriti:2013aqa}.}
This model has a canonical conjugate pair of three-index tensors as dynamical variables, and 
is formulated as a totally constrained system with first-class constraints, the algebra of which
resembles that of the ADM formalism \cite{Arnowitt:1960es} of general relativity.
A number of intriguing results have been obtained so far in this direction: 
there exists a formal continuum limit in which the first-class constraint algebra agrees with 
that of the ADM formalism \cite{Sasakura:2011sq}, 
the model is unique under 
some physically reasonable assumptions \cite{Sasakura:2012fb},
the constraints can consistently be quantized to form a first-class 
quantized constraint algebra with no anomaly \cite{Sasakura:2013wza},
locality is favored at least for $N=2$ in the physical wave-function 
satisfying the quantized constraints \cite{Sasakura:2013wza},
the $N=1$ case classically agrees with the mini-superspace 
approximation of general relativity \cite{Sasakura:2014gia},
and 
the model with $N=2$ has intimate relations with Ising model 
on random networks\cite{Sasakura:2014zwa,Sasakura:2014yoa}.

While the above results suggest that the canonical tensor model 
can be an interesting model of quantum gravity,
a major question on the validity of the model would be 
whether its quantum dynamics can produce an object like space-time.   
This question may be answered by studying the 
physical states (or physical wave-functions in explicit representations) 
which satisfy the quantized first-class constraints of the canonical tensor 
model.\footnote{In analogy with Hamilton formalism of general 
relativity, the quantized constraint equations and the physical wave-functions 
respectively correspond to the Wheeler-DeWitt equations and 
the Wheeler-DeWitt wave-functions\cite{DeWitt:1967yk}.}  
This is rather a difficult task due to the non-linear character of 
the quantized constraints and so far has been solved only in the case of 
$N=2$ \cite{Sasakura:2013wza}. 
On the other hand, the previous paper \cite{Sasakura:2014zwa} implies 
the intimate relations between the canonical tensor model and 
statistical systems on random networks\cite{Dorogovtsev,Sasakura:2014yoa}.
In fact, the present paper successfully constructs physical 
wave-functions satisfying the quantized constraints, 
general for arbitrary integer $N$, in terms of scale free integration of 
``grand-type'' partition functions of statistical systems on random networks (or rather random tensor networks as we will see).
We also obtain other physical wave-functions either by generalizing them for any $N$ or 
by explicitly solving the differential equations representing the constraints for $N=2,3$.
The following two properties of the solutions suggest 
physically interesting directions of future study.
One is that the wave-functions can have peaks at 
configurations where symmetries are enhanced.  
This is explicitly observed for $N=3$, and we expect this to be true also for higher $N$.
In future study, this feature may provide a clue 
as to why our universe is isotropic and homogenous 
at large scales. We also find that the physical wave-functions 
satisfying the constraints for $N=m$ with a cosmological constant 
can be obtained by restricting the domains of  
those for $N=m+1$ with no cosmological constant. 
This suggests that, in the canonical tensor model,
a cosmological constant is not an input parameter, but rather 
a part of dynamics. This would provide an interesting clue to the 
problem of the cosmological constant in future study.

This paper is organized as follows. In Section \ref{Introduction}, 
we recapitulate the canonical tensor model and its quantized 
constraints with a cosmological constant. A classification of physical 
states in terms of their kinematical and dynamical natures is introduced.
In Section \ref{n2}, we obtain the physical wave-functions satisfying 
the quantized constraints for $N=2$ generally with a cosmological constant
by explicitly writing down and solving the partial differential equations.
In Section \ref{n3}, we obtain physical wave-functions on an $S_3$ symmetric 
subspace of the configuration space in the case of $N=3$ generally 
with a cosmological constant.
We find that the wave-functions can have peaks at configurations 
with enhanced symmetries. In Section \ref{sec:ransol}, we construct 
physical wave-functions for general $N$ with no cosmological constant 
in terms of scale free integration of partition functions of statistical systems on random 
networks. In Section \ref{sec:simpsol}, we obtain physical wave-functions 
with simpler expressions by generalizing the result in Section \ref{sec:ransol}. 
In Section \ref{sec:explicitexample}, we explicitly evaluate the 
wave-functions in Section \ref{sec:simpsol} for $N=1, 2$
to check the validity of the expressions.
In Section \ref{sec:solPrep} and \ref{sec:solmatten}, 
we work in a momentum representation and obtain other physical wave-functions 
for the case of no cosmological constant and general $N$. 
We also obtain other types of physical wave-functions in 
terms of integration over matrix and tensor variables. In Section \ref{sec:WDWcosmo}, 
we obtain physical wave-functions for general $N$ with a cosmological constant 
by generalizing the results in Section \ref{sec:solPrep}. We find that 
these wave-functions for $N=m$ are just obtained by restricting 
the domains of the wave-functions for $N=m+1$ with no cosmological constant.  
A general theorem of this aspect is given in Section \ref{sec:generalcosmo}.
Finally, Section \ref{discussions} is devoted to summary and discussions.


\section{Canonical tensor model}
\label{Introduction}

The canonical tensor model has been introduced as a theory of 
dynamical fuzzy spaces \cite{Sasakura:2011sq, Sasakura:2012fb, 
Sasakura:2013gxg, Sasakura:2013wza}, 
aiming  to construct a quantum theory of gravity. 
First of all, the fuzzy space is an extended notion of space 
defined by a set of functions, $f_a\ (a=1,2,\cdots ,N)$, 
and the product of such functions:
\[
f_a \star f_b = M_{ab}{}^c f_c,
\label{fuzzy}
\]  
where $M_{ab}{}^c$ is a three-index tensor characterizing the fuzzy space. 
For instance, in the case of ordinary $d$-dimensional Euclidean space, 
(\ref{fuzzy}) is given by
\[
f_{z_1} \star f_{z_2} = \delta^d (z_1-z_2) f_{z_1}.
\]
Here $f_z$ can be expressed by a $d$-dimensional delta 
function, $f_z=\delta^d (x-z)$ where $x,z\in \mathbb{R}^d$.
This implies that points in Euclidean space correspond to functions, $f_z$'s.
As the three-index tensor is a delta function, distant points (functions) are independent. 
One can extend this notion of localized points to non-local ``fuzzy''  ones 
using a non-trivial three-index tensor, $M_{ab}{}^c$. 
Therefore, in general a fuzzy space is non-local and can be of any dimension.
In the canonical tensor model, in order to define the physics controlling the dynamics, 
two external conditions are imposed \cite{Sasakura:2011ma}. 
These are the reality conditions given by
\[
f^{\ast}_a = f_a, \ \ \ (f_a \star f_b)^{\ast} = f_b \star f_a,
\label{reality}
\]
where $\ast$ means complex conjugation, 
and the trace-like property of the inner product,
\[
\langle f_a | f_b \star f_c \rangle = \langle f_a \star f_b | f_c \rangle 
= \langle f_c \star f_a | f_b \rangle, 
\label{trace}
\] 
where the inner product, $\langle f_a | f_b \rangle$, has been chosen to 
be real, symmetric and bilinear. If one initially sets the inner product to be 
positive-definite, one can choose an orthonormal basis:
\[
\langle f_a | f_b \rangle = \delta_{ab},
\label{inner}
\] 
by a real linear transformation preserving the two conditions stated in
(\ref{reality}) and (\ref{trace}). This choice of orthonormal basis 
in (\ref{inner}) allows to rewrite (\ref{fuzzy}) in the following form:
\[
M_{abc} = \langle f_a \star f_b | f_c \rangle 
= M_{ab}{}^{d}\langle f_d | f_c \rangle.
\label{tensor}
\]
Using the three-index tensor $M_{abc}$ from (\ref{tensor}), it is 
possible to extract the dynamics of the fuzzy space under the 
imposed conditions stated in (\ref{reality}) and (\ref{trace}).
The inner product in (\ref{inner}) is $O(N)$ invariant, which 
appears in the transformations of $M$ as
\[
M'_{abc} = L_a{}^d L_b{}^e L_c{}^f M_{def}, \ \ \ L \in O(N). 
\label{on}
\]  
This symmetry serves as a kinematical symmetry of the canonical 
tensor model. The two conditions, (\ref{reality}) and (\ref{trace}), lead to 
the generalized hermiticity condition of the three-index tensor:  
\[
M_{abc} = M_{bca} = M_{cab} = M^{\ast}_{bac} = M^{\ast}_{acb} = M^{\ast}_{cba}. 
\label{hermite}
\]
The canonical tensor model stands on the position 
such that space-time would be a time evolution of the 
dynamical fuzzy space satisfying (\ref{reality}) and (\ref{trace}), 
or equivalently the generalized hermiticity condition, (\ref{hermite}).
This can be realized in analogy with the ADM formalism of general relativity \cite{Arnowitt:1960es}.
In the ADM formalism of general relativity, 
one considers the $4$-dimensional space-time as a time 
propagation (history) of the $3$-dimensional spatial hyper-surface, 
or in other words, parametrizes the metric in terms of 
non-dynamical fields, $N$, $N^i \ (i=1,2,3)$, 
and dynamical spatial metric, $h_{ij}$:
\[
d s^2 = -N^2 d t^2 + h_{ij} (N^i d t + d x^i)(N^j dt + d x^j).
\label{admm}
\] 
Then one converts the Einstein-Hilbert action into the 
Hamiltonian using the spatial metric, $h_{ij}$, 
and its conjugate momentum, $p^{ij}$, as phase-space variables. 
As a result of general covariance, the Hamiltonian 
can be written as a linear combination of constraints:
\[
H_{\text{ADM}}=\int d^3x \left[ N(x) \mathcal{H}(x) 
+ N^i(x) \mathcal{H}_i(x) \right],
\label{admh}
\]
where $\mathcal{H}$ and $\mathcal{H}_i$ are the generators of temporal 
and spatial diffeomorphism, respectively. They are constraints, classified 
as the first class following Dirac's theory of constrained systems. 
In fact, the constrains form a closed algebra. 
The canonical tensor model has been constructed in the same 
spirt of the ADM formalism, \textit{i.e.}, 
considering the three-index tensor, $M_{abc}$, and its 
conjugate momentum, $P_{abc}$, as dynamical phase-space variables, 
with the model being introduced as a totally constrained Hamiltonian system. 
One can construct the generators corresponding to the 
$O(N)$-kinematical symmetry, (\ref{on}), 
playing a similar role as the spatial diffeomorphism in general relativity, 
and remarkably one can uniquely determine the analogue of the 
temporal diffeomorphism of general relativity, under reasonable assumptions
requiring closed constraint algebra, cubic terms at most, invariance under the 
time-reversal symmetry and connectivity \cite{Sasakura:2012fb}.      

In this paper we are interested in quantizing the canonical tensor 
model started in \cite{Sasakura:2013wza}. 
In the quantum version, one replaces the dynamical variable $M_{abc}$ and its 
conjugate by corresponding operators $\hat{M}_{abc}$ and 
its conjugate partner, respectively. This will be explained 
later in more detail. Besides, we especially consider the 
minimal version of the canonical tensor model dubbed 
\textit{minimal model} \cite{Sasakura:2013gxg},
defined by a canonical conjugate pair of real and 
symmetric tensors, $(\hM_{abc},\hP_{abc})$, 
\[
\hM_{abc}=\hM_{abc}^\dagger=\hM_{bca}=\hM_{cab}=\hM_{bac}=\hM_{acb}=\hM_{bca}, \ \ \hbox{and similarly for }\hP_{abc},
\]
which satisfy the following commutation relations:
\[
[ \hM_{abc}, \hP_{def} ] 
=\frac{i}{6} \sum_{\sigma}
 \delta_{a\sigma (d)} \delta_{b \sigma (e)} \delta_{c \sigma (f)},
 \ \ 
 [ \hM_{abc}, \hM_{def} ]=[ \hP_{abc}, \hP_{def} ]=0, 
 \label{commutator}
\] 
where the summation over $\sigma$ implies that over all the permutations 
of $d,e,f$. As the minimal model has less degrees of freedom, 
its quantization is easier to perform and the algebra is more 
tractable compared to the non-minimal version of the canonical tensor model.
However the minimal model is still complicated 
enough to capture the non-trivial physics of the system.
It therefore becomes a good theoretical laboratory 
to gain expertise and maturity in the subject
(see, say \cite{Sasakura:2013wza}).
The quantum dynamics of the minimal model can be described through the 
Hamiltonian operator \cite{Sasakura:2013wza},
\[
\hat{H}=\hN_a \hat{\mathcal{H}}_a +\hN_{[ab]} \hat{\mathcal{J}}_{[ab]}, 
\label{qh}
\] 
where $\hN_a$ and $\hN_{[ab]}$ are Lagrange multipliers, and 
\[
&\hat{\mathcal{H}}_a = \frac{1}{2} \left( \hat{P}_{abc} \hat{P}_{bde} \hat{M}_{cde} 
- \lambda \hat{M}_{abb} + i \lambda_{H} \hat{P}_{abb}   \right),  
&\lambda_H = \frac{(N+2)(N+3)}{12}\label{hH} \\
&\hat{\mathcal{J}}_{[ab]} = \frac{1}{4} \left( \hat{P}_{acd} \hat{M}_{bcd} 
- \hat{P}_{bcd} \hat{M}_{acd}  \right) \label{hJ}.
\]
Here summing over repeated indices is implied, while
the symbol $[ab]$ refers to anti-symmetry of indices 
such that $\hJ_{[ab]}=-\hJ_{[ba]}$.
Ingredients appearing in (\ref{qh}) are explained in order: 
$\hJ_{[ab]}$ serve as the generator of the $O(N)$-kinematical symmetry, (\ref{on}), 
and $\hH_a$ is the generator of the symmetry analogous to the temporal 
diffeomorphism in general relativity; they are first class constraints 
and form a closed constraint algebra:
\[
&[\hH (\xi^1), \hH (\xi^2)] = \frac{i}{6}\hJ \left( [\hat{\xi^1}, \hat{\xi^2} ] + 2 \lambda [\xi^1, \xi^2] \right), \notag \\
&[\hJ (\eta), \hH (\xi)] = \frac{i}{6} \hH \left(\eta \xi \right), 
\label{eq:constraintalg}\\
&[\hJ (\eta^1), \hJ (\eta^2)] = \frac{i}{6} \hJ \left( [\eta^1,\eta^2] \right), \notag
\]
where $\hH(\xi)=\xi_a\hH_a$, $\hJ (\eta)=\eta_{[ab]}\hJ_{[ab]}$ and 
$\hat{\xi}_{ab}=\hP_{abc}\xi_c$; $[\ ,\ ]$ denotes the matrix commutator for $\hat \xi^i$,
and $[\xi^1,\xi^2]_{[ab]}=\xi^1_a \xi^2_b-\xi^2_a \xi^1_b$. 
Note that this is not a Lie algebra with structure constants, but has non-linear structures as on the right-hand side of the first line.  
Following the standard nomenclature, we call $\hJ_{[ab]}$ and $\hH_a$ 
\textit{momentum} and \textit{Hamiltonian constraints}, respectively. 
$\lambda_H$ is a real constant introduced by the operator 
ordering and has been fixed by imposing the hermiticity
of $\hH_a$, while $\lambda$ is a real undetermined constant 
which we call \textit{cosmological constant}.  
The last physical naming can be justified by its role in comparison 
with the mini-superspace approximation of 
general relativity \cite{Sasakura:2014gia}.
It is worth to recall that, in the case of hermite tensors \eq{hermite}, the cosmological constant term is prohibited by 
the consistency of the constraint algebra \cite{Sasakura:2012fb},
while this term is allowed for the minimal case as above.

In the case when $\lambda=0$, one can consistently incorporate the 
dilation generator to the Hamiltonian operator as
$\hat{H} \to \hat{H} +\hN\hD$, where $\hN$ is the 
Lagrange multiplier and $\hD$ in operator form is given by
\cite{Sasakura:2013gxg},
\[
\hD=\frac{1}{6} \left( \hP_{abc}\hM_{abc}+i\lambda_{D} \right). 
\label{hD}
\]
Here $\lambda_D$ is a constant introduced by the operator ordering, and 
can be fixed as 
\[
\lambda_D = \frac{N(N+1)(N+2)}{12}
\label{eq:vallambdaD}
\] 
by imposing the hermiticity of $\hD$. 
The constraint algebra remains closed even with the inclusion 
of the dilation generator, as can be seen for $\lambda=0$ from the following,
\[ 
&[\hD, \hH (\xi)] = \frac{i}{6}\hH (\xi), \notag \\
&[ \hD, \hJ(\eta) ] =0.
\]
However, we will not use this dilation generator as a genuine generator 
but merely as a mathematical tool for simplifying the process to find a 
solution to the constraint equations introduced below. 

In the following sections, we will find the physical states satisfying the constraints,
\[
\hH_a \Psi = \hJ_{[ab]} \Psi = 0.
\label{wdw}
\]
Note that here we do not impose the $\hat {\cal D}$ constraint to define the physical states. 
For later convenience, we introduce the $M$-representation of $\hP_{abc}$ and 
the $P$-representation of $\hM_{abc}$ as follows:
\[
&\hP_{abc} = -i D^M_{abc} = -i \Delta (abc) \frac{\partial}{\partial M_{abc}}, 
\ \ \ \hM_{abc} = iD^P_{abc} = i \Delta (abc) \frac{\partial}{\partial P_{abc}},
\notag \\
&\Delta (abc) = 
\left\{
\begin{array}{ll}
1, & \hbox{for }a=b=c,\\
\frac{1}{3}, &\hbox{for }a=b\ne c, \ b=c \ne a, \ c=a \ne b, \\
\frac{1}{6}, &\hbox{for } a\ne b, \ b \ne c, \ c \ne a,
\end{array}
\right.
\label{defd} 
\]  
where $M$ and $P$ are the eigenvalues of $\hM$ and $\hP$, respectively,
and $D^{M,P}_{abc}$ are rescaled partial differentials satisfying 
\[
D^M_{abc}M_{def} &= D^P_{abc} P_{def} =\frac{1}{6} \sum_{\sigma}
 \delta_{a\sigma (d)} \delta_{b \sigma (e)} \delta_{c \sigma (f)}.
\]

Finally, let us introduce a classification of the physical states in two types in the case with no cosmological constant, $\lambda=0$.
In this case, from \eq{hH}, the Hamiltonian constraints can be rewritten in a form,
\[
\hat {\cal H}_a=\frac{1}{2} \hat P_{abc} \hat {\cal J}_{(bc)},
\]
where 
\[
\hat {\cal J}_{(ab)}= \frac{1}{2} \left( \hat P_{acd} \hat M_{bcd}
+\hat P_{bcd} \hat M_{acd} \right) + i \lambda_{H} \delta_{ab}
\label{eq:J()def}
\]
with the round brackets of $(ab)$ symbolically representing the 
symmetric feature of the two indices. Then, the solutions to the 
constraints can be classified into the following two types,
\[
&\hbox{Kinematical: }\hat {\cal J}_{(ab)}\Psi=\hat {\cal J}_{[ab]}\Psi=0, \\
&\hbox{Dynamical: } \hat {\cal H}_{a}\Psi=\hat {\cal J}_{[ab]}\Psi=0,\ {}^\exists \hat {\cal J}_{(ab)}\Psi\neq 0,\
\]
The naming of ``kinematical" comes from the fact that the operators, 
$\hat {\cal J}_{(ab)}$ and $\hat {\cal J}_{[ab]}$, form a $\hbox{gl}(N)$ Lie algebra. 
This is a linear Lie algebra with structure constants, and such kinematical 
physical states would reflect only the kinematical characters rather than 
the dynamics of the canonical tensor model. 
A naive expectation is that physically interesting dynamics is caused 
by the non-linear features of the constraint algebra \eq{eq:constraintalg} with 
structure functions\footnote{It would be more appropriate to call them ``structure operators", since we are considering the quantized case.}, 
as in general relativity. 
From this viewpoint, the dynamical states would be of more importance. 
From \eq{eq:J()def}, a necessary condition for a kinematical state, 
which is a convenient criterion in subsequent analysis, is given by
\[
\bigl[ \hat P_{abc} \hat M_{abc} + i N \lambda_{H}  \bigr] \Psi=0.
\]
In the $P$ and $M$ representations, this criterion is respectively represented as 
\[
&\left(P_{abc} D^P_{abc} +N \lambda_{H}\right) \Psi(P)=0, \label{eq:criteriaP} \\
&\biggl[ M_{abc} D^M_{abc}+ \frac{(N-1)N(N+2)}{12} \biggr] \Psi(M)=0. 
\label{eq:criteriaM}
\]

\section{$N=2$ model}
\label{n2}

In this section we will consider the case of $N=2$ tensor model
both with and without a cosmological constant. In either case, we will explicitly 
solve the constraint equations (\ref{wdw}) to find the 
physical wave-function.

\subsection{Case without a cosmological constant}
\label{n2without}

In this subsection we consider the $N=2$ canonical tensor model with no 
cosmological constant, thereby setting $\lambda =0$. 
In this model, in the $P$-representation, there are $4$ independent 
variables, $\{P_{111}, P_{112}, P_{122}, P_{222} \}$.
On the other hand, the constraint equations \eq{wdw} give only three independent first 
order partial differential equations:
\[
\hH_1 \Psi (P) = \hH_2\Psi (P) = \hJ_{[12]} \Psi (P) =0 \, .
\label{wdwn2}
\] 
This implies that the most general solution to (\ref{wdwn2}) 
can be written with an arbitrary function of a single variable. 
In order to find this, we first consider a solution on a two-dimensional 
subspace, and then write it in an $O(2)$-invariant form
to extend it on the whole space. 

We start by introducing a $2$-dimensional subspace (or a gauge choice with respect to ${\cal J}_{[12]}$ and ${\cal D}$)
in the $P$-representation: 
\[
P_{111} =1, \ \ \ 
P_{112} =0, \ \ \ 
P_{122}=x_1, \ \ \ 
P_{222} =x_2,
\label{n2sub}
\]
where $x_1,x_2$ are the variables parameterizing the subspace.
In this $2$-dimensional subspace, 
the constraint equations (\ref{wdwn2}) are given by 
\[
\hH_1 \Psi &\propto \left[ 3 \frac{\partial}{\partial P_{111}} +x_1 (1+2x_1) 
\frac{\partial}{\partial x_1} +3x_1x_2 \frac{\partial}{\partial x_2} + 5(1+x_1) \right] \Psi =0, 
\label{n2h1} \\
\hH_2 \Psi &\propto \left[ x_1 (1+2x_1) \frac{\partial}{\partial P_{112}} 
+ 3x_1x_2 \frac{\partial}{\partial x_1} + 3(x_1{}^2 +x_2{}^2) \frac{\partial}{\partial x_2} + 5x_2 \right] \Psi =0, 
\label{n2h2} \\
\hJ_{[12]} \Psi &\propto \left[ (1-2x_1)\frac{\partial}{\partial P_{112}} 
-x_2 \frac{\partial}{\partial x_1} + 3x_1 \frac{\partial}{\partial x_2}  \right] \Psi =0 
\label{n2j} \, .
\]
By using (\ref{n2j}), we remove $\partial/\partial P_{112}$ from (\ref{n2h2}), 
which is a direction straying away from the subspace (\ref{n2sub}).
This gives us the following equation,
\[
\left[ 4x_1 x_2 (x_1 - 1) \frac{\partial}{\partial x_1} 
+ 3( 4x_1{}^3 + 2x_1 x_2{}^2 -x_2{}^2 ) \frac{\partial}{\partial x_2} 
+ 5x_2 (2x_1 -1) \right] \Psi =0.
\label{n2h2j}
\]
In fact, a solution to (\ref{n2h2j}) has been found in \cite{Sasakura:2013wza}
\footnote{
This is the exact solution of the $N=2$ canonical tensor model 
incorporating also the dilation constraint \eq{hD}.  
However, (\ref{n2withd}) of course satisfies (\ref{n2h2j}) in the present case. 
}:
\[
\Psi = \frac{\sqrt{ 4x_1{}^3 +x_2{}^2 }}{ x_1{}^2 ( x_1 -1 )^2 }.
\label{n2withd}
\] 
Then, the general solution to (\ref{n2h2j}) is given by
\[
\Psi = f(x_1,x_2)  \frac{\sqrt{ 4x_1{}^3 +x_2{}^2 }}{ x_1{}^2 ( x_1 -1 )^2 },
\] 
where $f(x_1,x_2)$ satisfies the homogeneous part of (\ref{n2h2j}), 
\textit{i.e.},
\[
\left[  4x_1 x_2 (x_1 - 1) \frac{\partial}{\partial x_1} 
+ 3( 4x_1{}^3 + 2x_1 x_2{}^2 -x_2{}^2 ) \frac{\partial}{\partial x_2} \right] f=0.
\]
This implies that $f$ is constant along the \textit{characteristics},
\[
\frac{dx_2}{dx_1} = \frac{ 3( 4x_1{}^3 + 2x_1 x_2{}^2 -x_2{}^2 ) }{ 4x_1 x_2 (x_1 -1) } \, .
\]
This can be solved to obtain,
\[
c_0 = \frac{ (4x_1{}^3 +x_2{}^2)^4  }{ x_1{}^6 ( 1-x_1 )^6 } \, ,
\]
where $c_0$ is a constant. Therefore, we get the following result,
\[
f(x_1 , x_2) = g \left( \frac{ (4x_1{}^3 +x_2{}^2)^4  }{ x_1{}^6 ( 1-x_1 )^6 } 
\right) \, ,
\]
where $g(z)$ is an arbitrary function of $z$. 
This solution can be written in an $O(2)$-invariant form.
As demonstrated in \cite{Sasakura:2013wza},
on the $2$-dimensional subspace (\ref{n2sub}), 
one can explicitly check that
\[
&A(P)= \epsilon_{ac} \epsilon_{bd} \epsilon_{eg} \epsilon_{fh} 
\epsilon_{e'g'} \epsilon_{f'h'} P_{aef} P_{bgh} P_{ce'f'} P_{dg'h'} =-2( 4x_1{}^3 +x_2{}^2), 
\label{eq:defAP}\\
&B(P)= P_{acd}P_{bde}P_{bef}P_{afc} - P_{acd}P_{bde}P_{aef}P_{bfc} =2 (1 -x_1 )^2 x_1{}^2, 
\label{eq:defBP}
\] 
where $\epsilon_{12}=-\epsilon_{21}=1,\, \epsilon_{11}=\epsilon_{22}=0$. 
Therefore, one can extend the solution above to an $O(2)$-invariant form:
\[
\Psi (P) = g \left( \frac{ A(P)^4 }{ B(P)^3 } \right) \frac{ \sqrt{A(P)} }{ B(P) }. 
\label{n2sol}
\]
In fact, one can explicitly check that (\ref{n2sol}) solves
(\ref{wdwn2}) on the whole space. 
Since the solution (\ref{n2sol}) contains an arbitrary function 
of a single variable, $g(z)$, it should be the most general 
solution to (\ref{wdwn2}). 

Let us classify the general solution \eq{n2sol} into the kinematical and dynamical solutions discussed in the last of Section \ref{Introduction}.
The necessary criterion \eq{eq:criteriaP} for a kinematical solution requires that a wave-function be a homogenous function of $P$ 
with degree $-N \lambda_H=-\frac{10}{3}$. This determines $g(z)=z^{-\frac{1}{3}}$, and therefore,
\[
\Psi_{kin}(P)=A(P)^{-\frac{5}{6}}.
\label{eq:psikin}
\]
In fact, one can explicitly check that, not only the necessary criterion, but also the full conditions for a 
kinematical state, $\hat {\cal J}_{(ab)} \Psi_{kin}=0$, are satisfied by \eq{eq:psikin}.  
Then, the other wave-functions than \eq{eq:psikin} are the dynamical ones.
This in turn implies that the potential singularities at $B(P)=0$ of the general wave-function \eq{n2sol} are purely of dynamical origin. 
As was argued in \cite{Sasakura:2013wza}, the configurations satisfying $B(P)=0$ are those of maximal locality. This would mean that
the non-linear character of the constraints is essentially important in the emergence of locality in the canonical tensor model.

The characteristics of the configurations satisfying $A(P)=0$, which are of kinematical origin as shown above, can be discussed 
in terms of the fuzzy space interpretation presented in Section \ref{Introduction}. 
Let us consider the following condition:
\[
v_a v_b P_{abc} =0,
\label{eq:vvP}
\]
where $v$ is a non-vanishing vector. Then, in the gauge \eq{n2sub}, one can easily show that 
such a non-vanishing $v$ exists, if and only if $4 x_1^3+x_2^2=0$, which exactly corresponds to $A(P)=0$.  
In the fuzzy space interpretation of the canonical tensor model explained in Section \ref{Introduction}, 
the ``points'' in a fuzzy space, $f_a$'s, are assumed to form an 
algebra\footnote{Here we do not care about which of $M$ or $P$ defines the fuzzy space algebra.
At present, we have no good argument to determine which choice is more proper.} ,
\[
f_a\star f_b=P_{abc}f_c.
\]
Thus the condition \eq{eq:vvP} implies that a ``point", $f = v_af_a$, satisfies
\[
f\star f =0.
\]
Therefore, $A(P)=0$ has the meaning that there exists a ``point'', $f$, with the property of a Grassmann number. 
We will see in Section \ref{n3without} that configurations with the same nature appear as potential singularities also
in the case of $N=3$.

\subsection{Case with a cosmological constant}
\label{n2with}

We now consider the $N=2$ tensor model with a cosmological constant. 
Since the corresponding constraint equations are more involved 
and finding the most general solution does not seem straightforward on first sight,
we will first solve them on a subspace invariant under an $S_2$ transformation. 
We will then extend the solution over the whole space by finding an $O(2)$ invariant expression.
The reason why we consider in this subsection such a symmetric subspace rather than \eq{n2sub} 
is that the equations are simplified because of the symmetry and
a similar procedure as below can also be applied to the case of $N=3$ as in Section \ref{n3}.

The subspace we consider is parameterized by 
\[
P_{111}=P_{222}=y_1, \ \ \ P_{112}=P_{221}=y_2,
\label{n2x1x2}
\]
which are the fixed points of 
the $S_2$ transformation permuting the index set, $\{1,2\}$.
The method of characteristics employed in Section~\ref{n2without} 
reduces the problem of solving the set of the first-order partial differential 
equations representing the constraints to solving the ordinary 
first-order differential equations along the flows generated by the constraints.
An infinitesimal variation of $\hP$ generated by the constraints is given by
\[
\delta \hP_{abc} = i \left[ \hP_{abc}, \ \xi_d \hH_d + \eta_{[de]} \hJ_{[de]}  \right] \, ,
\label{n2vari}
\]
where $\xi$ and $\eta$ are infinitesimal parameters. 
One can easily show that the infinitesimal variation (\ref{n2vari}) goes out of
the subspace (\ref{n2x1x2}), unless we require
\[
\xi_1 = \xi_2=\xi\, , \ \ \ \eta_{[ab]}=0 \, .
\]  
This requirement in turn determines the constraint equations on the $S_2$ subspace as
\[
\left( \hH_1 + \hH_2 \right) \Psi_{\lambda} = 0 \, ,
\label{n2wdwlam}
\]
where $\Psi_{\lambda}$ is an wave-function. By noting that 
\[
\frac{\partial}{\partial y_1} = \sum^{2}_{a=1} \frac{\partial}{\partial P_{aaa}}, 
\ \ \ 
\frac{\partial}{\partial y_2} = \sum^{2}_{ \substack{ a,b=1 \\ a \ne b}} \frac{\partial}{\partial P_{abb}},
\]
(\ref{n2wdwlam}) becomes
\[
\left[ 3 \left( y_1{}^2 +2y_2{}^2 + y_1 y_2 - \lambda \right) \frac{\partial}{\partial y_1} 
+ \left( 7y_2{}^2 + 5y_1 y_2 - \lambda  \right) \frac{\partial}{\partial y_2} 
+ 10 \left( y_1 + y_2 \right) \right] \Psi_{\lambda} =0.
\label{eq:partdifn2}
\]

From the result of Section \ref{n2without}, we know that the solution for $\lambda=0$ is given by \eq{n2sol} with
\[
A(y)&=-2(y_1 + 3y_2)( y_1 - y_2 )^3,  \\
B(y)&=\left[ 2\sqrt{2}(y_1-y_2) y_2 \right]^2,
\]
where $A(y),B(y)$ have been obtained by putting \eq{n2x1x2} to \eq{eq:defAP} and \eq{eq:defBP}, respectively.
In fact, $A(y), B(y)$ satisfy the following peculiar properties,
\[
{\cal O}_{0} A(y)& =12(y_1+y_2) A(y),
\label{eq:OAy} \\
{\cal O}_{0} B(y)& =16(y_1+y_2) B(y), 
\label{eq:OBy}
\]
where ${\cal O}_{0}={\cal O}_{\lambda=0}$ with ${\cal O}_{\lambda}$ defined by
the derivative part of \eq{eq:partdifn2},
\[
{\cal O}_{\lambda}=
3 \left( y_1{}^2 +2y_2{}^2 + y_1 y_2 - \lambda \right) \frac{\partial}{\partial y_1} 
+ \left( 7y_2{}^2 + 5y_1 y_2 - \lambda  \right) \frac{\partial}{\partial y_2}. 
\]
This means that \eq{eq:partdifn2} with $\lambda=0$ can be rewritten as a partial differential equation with variables $A,B$ rather than $y_1,y_2$, 
because, from the Leibniz rule,  
\[
\left[ {\cal O}_0+10 \left( y_1 + y_2 \right) \right] \Psi(y)&=\left[ \left({\cal O}_0 A \right)\frac{\partial}{\partial A}
+ \left({\cal O}_0 B \right)\frac{\partial}{\partial B}+10 \left( y_1 + y_2 \right) \right] 
\Psi(A,B) \CR
&=(y_1+y_2)\left[ 12 A \frac{\partial}{\partial A} + 16 B \frac{\partial}{\partial B} +10 \right] \Psi(A,B)=0.  
\]
When $A,B$ are regarded as variables, the expression \eq{n2sol} is indeed the most general solution to the simple partial differential equation
in the last line.
Therefore, if we keep the properties \eq{eq:OAy} and \eq{eq:OBy} for general $\lambda$, the solution
to \eq{eq:partdifn2} is simply given by the same expression as \eq{n2sol} with proper replacements of 
$A(y)$ and $B(y)$. Namely,
the solution is given by
\[
\Psi_{\lambda} (y) = g \left( \frac{ A_{\lambda} (y)^4}{ B_{\lambda} (y)^3 } \right) 
\frac{ \sqrt{A_{\lambda} (y) } }{B_{\lambda}(y)},
\label{n2glamsol}
\]
where $g(z)$ is an arbitrary function, and $A_\lambda(y),B_\lambda(y)$ must be determined through
\[
\frac{ {\cal O}_{\lambda} A_\lambda(y)}{A_\lambda (y)} = \frac{ {\cal O}_{0} A(y)}{A(y)}, \ \ 
\hbox{similarly for }B_\lambda(y).
\label{eq:OABylam}
\]
It does not seem always guaranteed that one can obtain the solutions of $A_\lambda(y), B_\lambda(y)$ 
to \eq{eq:OABylam} in simple expressions. 
But, rather miraculously, we obtain
\[
&A_{\lambda} (y) = -2(y_1 + 3y_2)( y_1 - y_2 )^3 
+ 4\lambda (y_1 + y_2)^2 -20 \lambda (y_1 -y_2)y_2 - \lambda^2, \label{n2a} \\
&B_{\lambda}(y) = \left[ 2\sqrt{2}(y_1-y_2) y_2 
- \frac{\sqrt{2}}{4}\lambda \right]^2 \label{n2b}.
\]
The reason why we get such simple expressions even for general $\lambda$ is probably related 
with what will be discussed in Section \ref{sec:generalcosmo}.

The solution in the whole space can simply be obtained by finding 
the $O(2)$-invariant expressions of $A_\lambda (y),B_\lambda (y)$:
\[
&\Psi_{\lambda} (P) = g \left( \frac{ A_{\lambda} (P)^4}{ B_{\lambda} (P)^3 } \right) 
\frac{ \sqrt{A_{\lambda} (P) } }{B_{\lambda}(P)},\\
&A_{\lambda}(P) = I_2(P) - 5 \sqrt{2} \lambda I_1(P) 
+ 2\lambda I_3 (P) - \lambda^2, \label{n2ap} \\
&B_{\lambda}(P) = \left[ I_1 (P) - \frac{\sqrt{2}}{4}\lambda \right]^2 \label{n2bp},
\]
where
\[
&I_1(P) = \frac{1}{\sqrt{2}} \epsilon_{i_1 i_2} 
\epsilon_{i_3 i_4} P_{i_1 i_3 j} P_{i_2 i_4 j}, \\
&I_2(P) =  \epsilon_{j_1l_1} \epsilon_{j_2l_2} 
\epsilon_{i_1i_2} \epsilon_{i_3i_4} P_{i_1i_3j_1} P_{i_2i_4j_2} 
\epsilon_{k_1k_2} \epsilon_{k_3k_4} P_{k_1k_3l_1} P_{k_2k_4l_2}, \\
&I_3(P) = P_{ijj} P_{ikk}. 
\]
In fact, one can explicitly check this is the solution in the whole space.

\section{$N=3$ model}
\label{n3}

In this section we extend our horizons further and explore the 
$N=3$ tensor model both with and without a cosmological 
constant. While, in this more complicated model, it is hard to 
find the general solution to the constraint equations (\ref{wdw}), 
it is possible to find the most general solution in a subspace 
satisfying an $S_3$ symmetry. 
The procedure is basically the same as that employed for $N=2$ in Section \ref{n2with}.
 
\subsection{Case with no cosmological constant}
\label{n3without}

Here we consider the $N=3$ minimal tensor model with no 
cosmological constant, and find the general solution to the 
constraint equations on a subspace invariant under an $S_3$ symmetry. 
The subspace is parametrized by 
\[
P_{aaa}=x_1, \ \ \ P_{abb}=x_2, \ \ \ P_{abc}=x_3,
\label{n3x1x2x3}
\]  
for any $a \ne b \ne c \ne a$,
which are the fixed points of the $S_3$ transformations permuting the index set $\{1,2,3 \}$. 
As in Section \ref{n2with}, 
one can show that 
there exists only one linear combination of the constraints
which generates a flow along the $S_3$ symmetric subspace \eq{n3x1x2x3},
and it is given by
\[
\hC^{(3)} = \sum^{3}_{a=1} \hH_a \, .
\label{n3const}
\]
On the subspace (\ref{n3x1x2x3}), 
the constraint (\ref{n3const}) is expressed as  
\[
\hC^{(3)} &= \frac{i}{2} \left( x_1{}^2 +2x_1x_2+2x_2(2x_2+x_3) \right) 
\frac{\partial}{\partial x_1} \notag \\
&\ \ \ +\frac{i}{6}\left( 5x_1x_2 +12x_2{}^2 +x_1x_3 + 7x_2x_3 +2x_3{}^2  \right) 
\frac{\partial}{\partial x_2} \notag \\
& \ \ \ +\frac{i}{2} \left( x_1x_3 + 4x_2 (x_2 +x_3) \right) \frac{\partial}{\partial x_3} 
+ \frac{3}{2}i \lambda_{H} (x_1 +2x_2) \, ,
\label{hC3}
\]
where $\lambda_H=\frac{5}{2}$, and we have used
\[
\frac{\partial}{\partial x_1} = \sum^{3}_{a=1} \frac{\partial}{\partial P_{aaa}}, 
\ \ \ 
\frac{\partial}{\partial x_2} = \sum^{3}_{ \substack{ a,b=1 \\ a \ne b}} 
\frac{\partial}{\partial P_{abb}}, 
\ \ \ 
\frac{\partial}{\partial x_3} = \frac{\partial}{\partial P_{123}}.
\]
In order to further simplify the process of finding a solution, 
let us virtually impose the dilation constraint (\ref{hD}):
\[
\hD= \frac{i}{6} \left( \sum^{3}_{i=1} x_i \frac{\partial}{\partial x_i} + \lambda_{D} \right),
\label{vhD}
\]
where $\lambda_{D}$ is supposed to be a free parameter for the purpose, rather than the fixed value in \eq{eq:vallambdaD}.

By assuming the two constraints, \eq{hC3} and \eq{vhD}, and $\lambda_D=0$ for the time being, 
we can eliminate $\partial/\partial x_1$ to obtain a constraint,
\[
\hC^{(3)}_v &= \frac{i}{6x_1} (2x_2+x_3)
\left\{  (x_1{}^2 + 3x_1x_2 -6x_2{}^2 + 2x_1 x_3 ) \frac{\partial}{\partial x_2}   
+  6x_2(x_1-x_3) \frac{\partial}{\partial x_3} \right\} \notag \\
& \ \ \  
+ \frac{3}{2}i \lambda_H (x_1+2x_2) \, .
\]
The method of characteristics applied to this partial differential equation, 
$\hC^{(3)}_v \Psi=0$, reduces the problem to solving 
\[
& ds = \frac{dx_2}{(2x_2 +x_3)(x_1{}^2 + 3x_1x_2 -6x_2{}^2 +2x_1x_3)} 
= \frac{dx_3}{6x_2(2x_2+x_3)(x_1-x_3)}, \label{ds} \\
& \frac{d\Psi}{\Psi} + 9\lambda_Hx_1(x_1+2x_2) ds=0 \, ,
\]
where $s$ parametrizes the trajectory of the flow. 
Note that $x_1$ is regarded as a constant in these equations. 
The trajectory equation derived from (\ref{ds}), 
\[
\frac{dx_2}{dx_3} = \frac{x_1{}^2 + 3x_1x_2 - 6x_2{}^2 +2x_1x_3}{6x_2(x_1-x_3)},
\]
tells us that the following combination remains unchanged along the trajectory:
\[
\frac{(-x_1+x_3)^3}{(-x_1+3x_2-2x_3)^2(x_1+6x_2+2x_3)}.
\] 
This gives a motivation to consider an ansatz:
\[
\Psi = \frac{(-x_1+x_3)^{c_1} (2x_2+x_3)^{c_2}}
{(-x_1 +3x_2 -2x_3)^{c_3} ( x_1+6x_2 +2x_3 )^{c_4}},
\label{n3vpsi}
\]
where $(2x_2+x_3)$ can be read off from (\ref{ds}), and $c_i$'s are numerical constants to be determined. 
Inserting (\ref{n3vpsi}) into the equations, $\hC^{(3)}\Psi = \hD \Psi =0$, 
one obtains
\[
c_2 &= \frac{3}{2} \left( \lambda_D -3\lambda_H  \right), \\
c_3 &= \frac{2}{3}c_1 + \lambda_D - \lambda_H, \\
c_4 &= \frac{1}{3}c_1 + \frac{1}{2} \left( 3 \lambda_D - 7\lambda_H \right).
\]  
Since $c_1$ and $\lambda_D$ can be taken arbitrary 
to satisfy the single constraint equation, $\hC^{(3)}\Psi=0$, 
the general form of $\Psi$ is determined to be
\[
\Psi &= (-x_1+3x_2-2x_3)^{\lambda_H} (2x_2 + x_3)^{-\frac{9}{2}\lambda_H} 
(x_1+6x_2+2x_3)^{\frac{7}{2}\lambda_H} \notag \\
& \ \ \ \times f 
\left( 
\frac{(2x_2+x_3)^3}{( -x_1+3x_2-2x_3 )^2(x_1+6x_2+2x_3)^3},
\frac{(-x_1+x_3)^3}{(-x_1+3x_2-2x_3)^2(x_1+6x_2+2x_3)}
\right),
\label{n3solwoc}
\]  
where $f$ is an arbitrary function with two arguments. 
It is clear from number counting that (\ref{n3solwoc}) gives the 
most general solution to the constraint equation, $\hC^{(3)} \Psi =0$.


Let us study peaks of the wave-function in (\ref{n3vpsi}). 
Since we cannot uniquely determine the coefficients,
$c_i$, 
here we consider possible peaks of the wave 
function (\ref{n3vpsi}) assuming signs of the coefficients.    
Firstly, if $c_3>0$, then (\ref{n3vpsi}) diverges when $-x_1+3x_2-2x_3=0$. 
This means that the configurations satisfying this equation are enhanced. 
In fact, it can be shown that, if $-x_1+3x_2-2x_3=0$, 
$P_{abc}$ is invariant under an infinitesimal SO($2$)  transformation:
\[
t_{aa'}P_{a'bc} + t_{bb'}P_{ab'c} + t_{cc'}P_{abc'} =0, 
\]
where
\[
t=
\begin{pmatrix}
0&1&-1\\
-1&0&1\\
1&-1&0
\end{pmatrix}.
\]
Therefore, the configurations with the SO($2$) symmetry are enhanced in the wave-function. 

Secondly, if $c_4>0$, then the wave-function (\ref{n3solwc}) 
diverges when $x_1+6x_2+2x_3=0$. 
This configuration satisfies
\[
v_a v_b P_{abc} =x_1+6x_2+2x_3=0,
\label{grassmann}
\]
where
\[
v=(1,1,1).
\]
This is the same situation discussed for the case of $N=2$ in the last paragraph of Section \ref{n2without}.
Therefore, the condition, $x_1+6x_2+2x_3=0$, has the meaning that 
there exists a ``point'', $f=v_a f_a$, with the property of a Grassmann number. 

A more interesting possibility arises for the configurations which simultaneously satisfy 
the two conditions above,  $-x_1+3x_2-2x_3=x_1+6x_2+2x_3=0$.
They are the configurations satisfying
\[
x_2=0, \ \ \ x_3=-\frac{x_1}{2}.
\]
What is interesting on these points is that, in addition to the SO($2$) symmetry mentioned above, 
$P_{abc}$ is invariant under an infinitesimal boost transformation, 
\[
k_{aa'}P_{a'bc} +k_{bb'}P_{ab'c} +k_{cc'}P_{abc'}=0,
\]
where
\[
k=
\begin{pmatrix}
0&1&1\\
1&0&1\\
1&1&0
\end{pmatrix}.
\]
Therefore, these configurations are invariant under a Lorentz group. 
In the present $N=3$ case, the generators commute with each other, $[k,t]=0$, and 
may not be of much interest. However, we expect enhancement of non-abelian Lorentz groups to occur for larger $N$.  

We could not find similar kinematical characterization as above for the other possible peaks at $2x_2+x_3=0$ or $-x_1+x_3=0$.
Probably, these are more rooted in dynamics, similarly to $B(P)=0$ for $N=2$ as discussed in Section \ref{n2without}.

\subsection{Case with a cosmological constant}
\label{n3with}

In this subsection, 
we generalize the solution on the $S_3$ symmetric subspace obtained in Section \ref{n3without} 
to include the cosmological constant $\lambda$
by applying the method employed in Section \ref{n2with} for the case of $N=2$.
With $\lambda$, the constraint \eq{hC3} is replaced by 
\[
\hC^{(3)}_{\lambda} &= \sum^{3}_{a=1} \hH_a \notag \\
&= \frac{i}{2} \left( -\lambda + x_1{}^2 + 2x_1x_2 + 2x_2 (2x_2 +x_3)  \right) 
\frac{\partial}{\partial x_1}  \notag \\  
& \ \ \  + \frac{i}{6} \left( -\lambda + 5x_1x_2 +12x_2{}^2 
+ x_1 x_3 +7x_2x_3 +2x_3{}^2  \right) \frac{\partial}{\partial x_2}  \notag \\
& \ \ \  + \frac{i}{2} \left( x_1x_3 + 4x_2 (x_2 +x_3) \right) \frac{\partial}{\partial x_3} 
+ \frac{3}{2}i \lambda_H (x_1+2x_2) \, .
\label{n3z3lam}
\] 
As in Section \ref{n2with}, let us take the derivative part of the constraint \eq{n3z3lam} as
\[
{\cal O}^{(3)}_{\lambda} &= \frac{i}{2} \left( -\lambda + x_1{}^2 + 2x_1x_2 + 2x_2 (2x_2 +x_3)  \right) 
\frac{\partial}{\partial x_1}  \notag \\  
& \ \ \  + \frac{i}{6} \left( -\lambda + 5x_1x_2 +12x_2{}^2 + x_1 x_3 +7x_2x_3 +2x_3{}^2  \right) 
\frac{\partial}{\partial x_2}  \notag \\
& \ \ \  + \frac{i}{2} \left( x_1x_3 + 4x_2 (x_2 +x_3) \right) \frac{\partial}{\partial x_3}. 
\label{n3D}
\]

The solutions (\ref{n3solwoc}) for $\lambda =0$ are expressed in terms of the following linear 
combinations of $x_i$'s:
\[
E_1 &= -x_1+3x_2 -2x_3, \\
e_2 &= 2x_2 +x_3, \\
e_3 &= x_1 + 6x_2 +2x_3, \\
e_4 &= -x_1 +x_3 \, .
\]
In the same way as the $N=2$ case, one would be able to construct solutions in the case of $\lambda \ne 0$ 
by considering corrections to these $E_1,e_{2,3,4}$ so that they satisfy the requirement corresponding to \eq{eq:OABylam}.
In fact, $E_1$ satisfies
\[
\frac{{\cal O}^{(3)}_{\lambda} E_1}{E_1} = \frac{i}{2} (x_1 -x_3).
\label{n3de1}
\]
The crucial point to note here is that the right-hand side of (\ref{n3de1})
is independent of $\lambda$, and therefore $E_1$ meets the requirement without any corrections.

Let us next consider $e_3$. In this case, 
\[
{\cal O}^{(3)}_{\lambda}e_3 = \frac{i}{2} (x_1+6x_2+2x_3)e_3 - \frac{3}{2}i \lambda,
\]
which requires us to make corrections to $e_3$. Since 
the dimension of $\lambda$ is that of $x_i{}^2$, 
$e_3$ is not appropriate to consider corrections in perturbation of $\lambda$. 
Instead, we would be able to consider $e_3{}^2$ and its correction as
\[
E_3 = e_3{}^2 + \gamma  \lambda,
\label{E3}
\]
where $\gamma$ is a numerical constant to be determined. 
Applying (\ref{n3D}) on $E_3$, one obtains
\[
{\cal O}^{(3)}_{\lambda}E_3 = i(x_1+6x_2+2x_3) \left( e_3{}^2 -3\lambda \right).
\]
Therefore, if one takes $\gamma=-3$ in (\ref{E3}), then
\[
\frac{{\cal O}^{(3)}_{\lambda}E_3}{E_3} = i(x_1+6x_2+2x_3). 
\]
The righthand side does not contain $\lambda$, and therefore $E_3$ satisfies the requirement. 
In a similar manner, 
we can find suitable combinations of variables as
\[
E_2&=e_2e_4 + \frac{\lambda}{4}, \\
E_4&=e_4{}^3 + \frac{3}{4}\lambda (2x_1-6x_2-5x_3).
\]
Thus, using the new set of variables, 
$\{E_1, E_2, E_3, E_4 \}$, one can generalize (\ref{n3solwoc}) to 
\[
\Psi_{\lambda} = E_1{}^{\lambda_H}E_2{}^{-\frac{9}{2}\lambda_H} 
E_3{}^{\frac{7}{4}\lambda_H} E_4{}^{\frac{3}{2}\lambda_H} 
f \left( \frac{E_2{}^3}{E_1{}^4E_3{}^2}, \frac{E_4{}^2}{E_1{}^4E_3}  \right),
\label{n3solwc}
\]
where $f(z_1,z_2)$ is an arbitrary function. 
In fact, one can explicitly check that (\ref{n3solwc}) satisfies
\[
\hC^{(3)}_{\lambda}\Psi_{\lambda} =0.
\]

As shown above for $N=3$,
we have again found that, similar to the case of $N=2$, the inclusion of the cosmological constant can be implemented by making 
some corrections to the solutions with no cosmological constant. 
This suggests that the inclusion of the cosmological constant does not change the essential structure of the
theory. In fact, we will see in Section \ref{sec:generalcosmo} 
that the cases with a cosmological constant can be treated by considering the cases with no cosmological constant
for increased $N$.

\section{Solutions in terms of statistical systems on random networks}
\label{sec:ransol}

So far we have studied the solutions of the constraint 
equations of the canonical tensor model for the simple 
cases of $N=2,3$ by explicitly writing down and solving the 
sets of the partial differential equations representing the constraints.
While this procedure can work for small $N$'s,
it is obvious that 
the situation becomes much more complicated as soon as one moves to higher-$N$ 
tensor model, and one quickly notices the limitations of the 
methodologies used in studying the $N=2,3$ cases.
In the present and following sections,
instead of giving general explicit solutions for particular $N$'s,  
we will present some series of solutions valid for general $N$, by using 
the partition functions of statistical systems on random networks and some variants. 

In \cite{Sasakura:2014zwa}, it was argued that the $N=2$ canonical tensor 
model is intimately related to the Ising model on random networks
of trivalent vertices. In fact, it has been shown that the phase structure of the Ising model on random networks 
can be derived from the Hamiltonian vector flow of the canonical tensor model for $N=2$, if the flow is regarded 
as the renormalization group flow of the Ising model. 
The present and following sections will give further relations between 
the canonical tensor model and statistical systems on random networks
by obtaining the physical wave-functions solving the constraints 
from the perspective of statistical systems on random networks.
 
Let us start with the following form of the ``grand-type" partition function of statistical systems 
on random networks \cite{Sasakura:2014zwa},
\[
Z(k,M,{\cal C}_\phi)= \int_{\cal C_\phi} d\phi\  e^{-k \phi^2 +M \phi^3},
\label{eq:zkmc}
\]
where we have used shorthand notations,  
\[
&d\phi\equiv \prod_{a=1}^N {\rm d}\phi_a, \CR
&\phi^2\equiv \phi_a \phi_a, \CR
&M \phi^3\equiv M_{abc}\phi_a\phi_b\phi_c.
\]
Here $k$ and $M_{abc}\ (a,b,c=1,2,\cdots,N)$ are a numerical variable and 
a three-index symmetric tensor, respectively. ${\cal C}_\phi$ denotes the 
domain of integration over $\phi_a$'s, which can generally take complex values.

As discussed in \cite{Sasakura:2014zwa}, 
to make relations with statistical systems on random networks,
$k$ is assumed to be a positive real number, and the integration contour, 
${\cal C}_\phi$, is taken, for instance among various 
allowed possibilities, as 
\[
\phi_a =e^{\frac{\pi i }{6}} r_a,
\label{eq:intdomain}
\]
where $r_a$ run from $-\infty$ to $\infty$ on the real axis.
By expanding the integrand of \eq{eq:zkmc} in $M$,
one obtains an asymptotic expansion in $M$ as 
\[
Z(k,M,{\cal C}_\phi)\simeq\sum_{n=0}^\infty Z_n, \ \ \   
Z_n= \frac{1}{n!} \int_{-\infty}^{\infty} d\phi \ (M \phi^3)^n e^{-k \phi^2}.
\label{eq:pertexpn}
\]
Each $Z_n$ gives the partition function of a statistical 
system on random networks of $n$ trivalent vertices \cite{Sasakura:2014zwa, Sasakura:2014yoa}.
In this paper, we do not generally assume the positivity of $k$ or
the integration contour like \eq{eq:intdomain}, since our interest is in the solutions to the 
constraint equations rather than such statistical systems. 
As we will see shortly, the necessary properties are the well-definedness of the integral \eq{eq:zkmc} 
in the space of $M$ except for possible singularities, and  the validity 
of the partial integrations over $\phi_a$'s which will be performed in the rest of this paper. 

Firstly, one can show that the partition function \eq{eq:pertexpn} satisfies the 
momentum constraints (\ref{hJ}), because
\[
\hat {\cal J}_{[ab]} Z(k,M,{\cal C}_\phi)&\propto (M_{acd} D^M_{bcd}-M_{bcd} D^M_{acd})Z(k,M,{\cal C}_\phi) \CR
&=\int_{\cal C_\phi} d\phi\  (M_{acd} \phi_b \phi_c \phi_d -M_{bcd} 
\phi_a \phi_c \phi_d )e^{-k \phi^2+M\phi^3} \CR
&=\frac{1}{3} \int_{\cal C_\phi} d\phi\  e^{-k \phi^2}
\left( \phi_b D^{\phi}_a  -\phi_a D^\phi_b \right)e^{M\phi^3}  \CR
&=-\frac{1}{3} \int_{\cal C_\phi} d\phi\  e^{M\phi^3} 
\left[D^{\phi}_a\left( \phi_b e^{-k \phi^2}\right) - D^\phi_b \left(\phi_a e^{-k \phi^2}\right)
\right] \CR
&=0 \, , 
\]
where we have used a short-hand notation,
\[
D^\phi_a\equiv \frac{\partial}{\partial \phi_a} \, ,
\]
and have assumed the validity of the partial integrations over $\phi_a$'s.
This is justified, if ${\cal C}_{\phi}$ is a closed curve,
or if the integrand damps rapidly enough in the case 
that ${\cal C}_{\phi}$ extends to infinity. 
In the rest of this paper, without being explicitly mentioned, 
we simply assume that all the partial integrations (not only 
over $\phi_a$'s but also over some other variables) performed in due course be valid. 

Next, let us consider the Hamiltonian constraints. From \eq{hH}, the corresponding partial differential equations 
for $\lambda=0$ in the $M$-representation are given by
\[
\left( M_{bde} D^M_{abc} D^M_{cde}+\lambda_H D^M_{abb} \right)\psi (M)=0.
\label{eq:partialinMlamzero}
\]
As for the first term, we obtain
\[
M_{bde} D^M_{abc} D^M_{cde} Z(k,M,{\cal C}_\phi) &=\int_{{\cal C}_\phi} d\phi\ 
M_{bde} \phi_a \phi_b \phi_c \phi_c \phi_d \phi_e 
e^{-k \phi^2 +M \phi^3} \CR
&= \int_{{\cal C}_\phi}  d\phi\ \phi_a \phi^2 M_{bde} 
\phi_b\phi_d \phi_e e^{-k \phi^2 +M \phi^3} \CR 
&=\frac{1}{3}\int_{{\cal C}_\phi}  d\phi\ \phi_a \phi^2 e^{-k \phi^2}  \phi_b D^\phi_b 
e^{M\phi^3} \CR
&=-\frac{1}{3} \int_{{\cal C}_\phi}  d\phi\ e^{M\phi^3}
D^\phi_b \left( \phi_a \phi_b \phi^2 e^{-k \phi^2} \right) \CR
&=-\int_{{\cal C}_\phi}  d\phi\ \phi_a \phi^2 
\left(1+\frac{N}{3}-\frac{2 k}{3} \phi^2 \right) e^{-k \phi^2+M\phi^3} \CR
&=-\int_{{\cal C}_\phi}  d\phi\ \phi_a \phi^2 
\left(1+\frac{N}{3}+\frac{2 }{3} k\frac{\partial}{\partial k} \right) e^{-k \phi^2+M\phi^3} .
\label{eq:hamfirst}
\]
As for the second term of \eq{eq:partialinMlamzero}, we obtain
\[
D^M_{abb} Z(k,M,{\cal C}_\phi) =\int_{{\cal C}_\phi}  
d\phi\ \phi_a \phi^2 e^{-k \phi^2+M\phi^3}. 
\label{eq:hamsecond}
\]

Because of the derivative term with respect to $k$ in the last line of \eq{eq:hamfirst}, 
the partition function itself does not satisfy the Hamiltonian constraints. 
This can be remedied by introducing a function $g(k)$ and the integration over $k$. To see this, let us define
\[
\psi_{{\cal C}_{k,\phi}}(M) =\int_{{\cal C}_k}\,dk \,g(k)\, Z(k,M,{\cal C}_\phi)
=\int_{{\cal C}_{k,\phi} }\,dkd\phi \,g(k)e^{-k \phi^2+M\phi^3}.
\label{eq:psirep}
\] 
Then 
\[
&\left( M_{bde} D^M_{abc} D^M_{cde}+\lambda_H D^M_{abb} \right)\psi_{{\cal C}_{k,\phi}} (M) \CR
&\hspace{2cm} = \int_{{\cal C}_{k,\phi}} dkd\phi\ g(k)\, \phi_a \phi^2 
\left[- \left(1+\frac{N}{3}+\frac{2 }{3} k\frac{\partial}{\partial k} \right)
+\lambda_{H} \right]
e^{-k \phi^2+M\phi^3} \\
&\hspace{2cm}
= \frac{2}{3} \int_{{\cal C}_{k,\phi}} dkd\phi\ \phi_a \phi^2 
\left( k g'(k)+\alpha_N g(k) \right)
e^{-k \phi^2+M\phi^3} ,
\]
where we have performed a partial integration over $k$ under the assumption of its validity,
and 
\[
\alpha_N=\frac{3  \lambda_H-N-1}{2}=\frac{N^2+N+2}{8}.
\label{eq:alphaN}
\]
Therefore the Hamiltonian constraints are satisfied, if we take
\[
g(k)=k^{-\alpha_{N}}.
\label{eq:gk}
\]
By considering various choices of the integration domain, ${\cal C}_{k,\phi}$, 
one can obtain a number of independent solutions.

Physics behind the present solution would be given as follows. 
As discussed in the previous paper \cite{Sasakura:2014zwa}, 
a flow generated by a scale-free $O(N)$-invariant linear combination of the Hamiltonian 
constraints of the canonical tensor model could be interpreted 
as a renormalization group flow of statistical systems on random 
networks.\footnote{Strictly speaking, this was discussed only 
for $N=2$ corresponding to the Ising model.}  
A renormalization group procedure, if existed,  on random networks would 
change the number of vertices of networks, that is a natural
analogy of a block spin transformation on a regular lattice.
Therefore, since the parameter $k$ controls the relative 
weights among networks of $n$ trivalent vertices 
as $\propto k^{-\frac{3n}{2}}$, a renormalization group 
procedure should generate a flow in $k$. This would be in accordance with
the appearance of a derivative term of $k$ in the last line of \eq{eq:hamfirst}, 
and it must be compensated by introducing $g(k)$ to make 
an invariant under the flow.

Here let us check the type of the solution $\psi_{{\cal C}_{k,\phi}}$.
We obtain
\[
M_{acd} D^M_{bcd} \psi_{{\cal C}_{k,\phi}}(M)  &= \int_{{\cal C}_{k,\phi}} 
dk d\phi\ g(k) \ M_{acd} \phi_b \phi_c \phi_d \ e^{-k \phi^2+M \phi^3} \CR
&=\frac{1}{3} \int_{{\cal C}_{k,\phi}} dk d\phi\ g(k) e^{-k \phi^2}    
\phi_b D^\phi_a e^{M \phi^3}  \CR
&=- \frac{1}{3} \int_{{\cal C}_{k,\phi}} dk d\phi\ g(k) \  e^{M \phi^3} 
D^\phi_a \left( \phi_b \  e^{-k \phi^2}\right)  \CR
&=- \frac{1}{3} \int_{{\cal C}_{k,\phi}} dk d\phi\ g(k) 
\left( \delta_{ab}-2 k \phi_a \phi_b \right) e^{-k \phi^2+M \phi^3}.
\label{eq:applyj}
\]
Here the second term in the last line cannot be concluded to be proportional to $\delta_{ab}$, 
and therefore $\hat {\cal J}_{(ab)} \psi_{{\cal C}_{k,\phi}}(M)\neq 0$ (See \eq{eq:J()def} for $\hat {\cal J}_{(ab)}$) is expected in general.
This means that  $\psi_{{\cal C}_{k,\phi}}$ is a dynamical solution in general.
However, there exists a delicate issue for $N=1,2$ as follows.
By performing similar computations as \eq{eq:applyj}, we obtain
 \[
 \left( M_{abc} D^M_{abc}+\lambda_H-1\right)  \psi_{{\cal C}_{k,\phi}}(M)=0.
 \label{eq:ApplyDtoCkpm}
 \]
Then one finds that, for $N=1,2$, $\psi_{{\cal C}_{k,\phi}}(M)$ satisfies the necessary condition \eq{eq:criteriaM} to be kinematical.
For $N=1,2$, we cannot ignore the possibility that $\psi_{{\cal C}_{k,\phi}}(M)$ may become kinematical.
For $N\geq 3$, since the necessary condition is violated, the solution is definitely dynamical.

%
%
%
%
%
%

Finally, we would like to comment on the partition function (\ref{eq:zkmc}) from a mathematical view point.
Let us start with a generalization of the Airy function:
\[
\text{Ai} \left[  j,M ,{\cal C}_\phi \right] 
= \int_{\mathcal{C}_{\phi}} d\phi \ e^{-j_a\phi_a+M_{abc}\phi_a\phi_b\phi_c},
\label{generalairy}
\]
where $j$ is a vector. 
(\ref{generalairy}) satisfies the following generalization of Airy's differential equation:
\[
\left( 3 M_{abc} \frac{\partial}{\partial j_b} \frac{\partial}{\partial j_c} - j_a \right) \text{Ai} \left[  j,M ,{\cal C}_\phi \right] =0,
 \]
which can be shown by partial integrations over $\phi$.
The generalized Airy function can be related to the partition function \eq{eq:zkmc} in the following manner.
Firstly, one can remove the linear term in the exponent in (\ref{generalairy})
by a shift,
\[
\phi_a \to \phi_a+w_a,
\]
where $w$ is a vector satisfying 
\[
j_a-3M_{abc}w_bw_c=0.
\]
Then there appears a quadratic term in $\phi$, $3 M_{abc} w_c \phi_a \phi_b$, as well as a constant term, $-j_a w_a+M_{abc}w_a w_b w_c$.
The quadratic term can be diagonalized and normalized by a linear transformation, 
\[
\phi_a \to \phi_b R_{ba}, 
\]
with a complex matrix $R$, if the quadratic term is not singular.
Then, (\ref{generalairy}) can be transformed  to a form,
\[
\text{Ai} \left[  j,M ,{\cal C}_\phi \right] =|R| e^{ -j w+M w^3} \int_{\mathcal{C}'_{\phi}} d\phi \  e^{-k\phi^2+M'\phi^3} 
=|R|e^{ -j w+M w^3}\, Z(k,M',{\cal C}'_\phi),
\]
where 
\[
M'_{abc}=R_{aa'}R_{bb'}R_{cc'}M_{a'b'c'},
\]
$k$ is a number, and $|R|$ denotes the determinant of $R$.
Therefore the mathematical properties of the generalized Airy functions can be related to 
those of the partition function. This observation would be useful in future study.

\section{Simpler solutions}
\label{sec:simpsol}

The solution found in Section \ref{sec:ransol} has a direct connection 
with statistical systems on random networks, 
and would therefore be interesting from physical viewpoints. 
On the other hand, if we set aside the physical interpretation, 
the solution \eq{eq:psirep} would be simplified by replacing 
$\int dk\, g(k) e^{-k \phi^2}$ with a function of $\phi^2$.
Thus, in this section, let us assume a form,
\[
\psi_{{\cal C}_\phi,h}(M)=\int_{{\cal C}_\phi} d\phi\ f(\phi^2) h(M \phi^3),
\label{eq:psimfh}
\]
where $f$ and $h$ are functions of the shown arguments, and solve the constraints.
$f$ will be determined shortly.

Let us first check the momentum constraints. 
We obtain
\[
\left( M_{acd} D^M_{bcd}-M_{bcd}D^M_{acd} \right) 
\psi_{{\cal C}_\phi,h}(M) &= \int_{{\cal C}_\phi} d\phi\ f(\phi^2) 
(M_{acd} \phi_b \phi_c \phi_d - M_{bcd} \phi_a \phi_c \phi_d ) h'(M \phi^3) \CR
&= \frac{1}{3} \int_{{\cal C}_\phi} d\phi\ f(\phi^2) 
( \phi_b D^\phi_a - \phi_a D^\phi_b) h(M \phi^3) \CR
&=-\frac{1}{3} \int_{{\cal C}_\phi} d\phi\ h(M \phi^3) 
\left( D^\phi_a(\phi_b f(\phi^2))-  D^\phi_b(\phi_a  f(\phi^2)) \right)  \CR
&=0 \, .
\]
As for the Hamiltonian constraints, we first obtain
\[
M_{bde} D^M_{abc} D^M_{cde} \int_{{\cal C}_\phi}  d\phi\ f(\phi^2) h(M \phi^3)
&=\int_{{\cal C}_\phi} d\phi\ M_{bde} \phi_a \phi_b \phi_c \phi_c \phi_d \phi_e f(\phi^2)
h''(M \phi^3) \CR
&=\int_{{\cal C}_\phi}  d\phi\ f(\phi^2) \phi_a \phi_b \phi^2 
\frac{1}{3} D^\phi_b  h'(M\phi^3) \CR
&=-\frac{1}{3} \int_{{\cal C}_\phi}  d\phi\ h'(M\phi^3)D^\phi_b
\left( f(\phi^2) \phi_a \phi_b \phi^2\right) \CR
&=-\frac{1}{3} \int_{{\cal C}_\phi} d\phi\ \phi_a \left( (N+1) \tilde f (\phi^2) 
+ 2 \phi^2 \tilde f'(\phi^2) \right) h'(M\phi^3), 
\label{eq:MDDsimpwave}
\]
where $\tilde f(x) = xf(x)$.
Similarly,
\[
D^M_{abb} \int_{{\cal C}_\phi}  d\phi\ f(\phi^2) h(M \phi^3)
&=\int_{{\cal C}_\phi}  d\phi \ f(\phi^2) \phi_a \phi_b \phi_b h'(M\phi^3), \CR
&=\int_{{\cal C}_\phi}  d\phi\ \phi_a \tilde f(\phi^2) h'(M\phi^3).
\]
Therefore,
\[
(M_{bde} D^M_{abc} D^M_{cde}+\lambda_H D^M_{abb})
\psi_{h,{\cal C}_\phi}(M) 
=\frac{2}{3} \int_{{\cal C}_\phi}  d\phi\ \phi_a 
\left( \alpha_N \tilde f (\phi^2) - \phi^2 \tilde f'(\phi^2) \right) h'(M\phi^3),
\]
where $\alpha_N$ was defined in \eq{eq:alphaN}.
Thus, by putting 
\[
\tilde f (x)= x^{\alpha_N},
\]
a solution to the constraint equations can be obtained as  
\[
\psi_{{\cal C}_\phi,h}(M)=\int_{{\cal C}_\phi} d\phi \ (\phi^2)^{\beta_N} h(M\phi^3),
\label{eq:simpwave}
\]
where $\beta_N=\alpha_N-1=\frac{(N-2)(N+3)}{8}$.
One would be able to obtain a number of independent solutions 
by considering various ${\cal C}_\phi$ and $h$. 

One can obtain another kind of solutions by determining 
$h$ instead of $f$ in \eq{eq:psimfh}. In this case, we have
\[
&\left( M_{cde}D^M_{abc}D^M_{bde}+\lambda_H D^M_{abb} \right)
\int_{{\cal C}_\phi} d\phi\ f(\phi^2) h(M \phi^3) \CR
& \hspace{3cm}=\int_{{\cal C}_\phi} d\phi\ \phi_a \phi^2 f(\phi^2)
\left( M \phi^3 h''(M\phi^3)+\lambda_H h'(M\phi^3)\right). 
\]
This vanishes, if 
\[
h(x)=x^{-\gamma_N}+\hbox{const.},
\]
where $\gamma_N=\lambda_H-1=\frac{(N-1)(N+6)}{12}$.
The constant term is irrelevant, since it does not produce any dependence on $M$. 
Thus a non-tivial solution to the constraints is given by 
\[
\psi_{{\cal C}_\phi,f}(M)=\int_{{\cal C}_\phi} d\phi\ (M\phi^3)^{-\gamma_N} f(\phi^2).
\label{eq:psifc}
\]
Various choices of ${\cal C}_\phi,f$ will provide a number of independent solutions.

For instance, by considering $f(x)=e^{-x}$ in \eq{eq:psifc}, we obtain 
a similar expression as $Z_n$ in \eq{eq:pertexpn}.
Therefore, in the large-$N$ limit, one would be able to employ a 
similar saddle-point method used in the analysis
of the thermodynamic limit of statistical systems on random networks 
in \cite{Sasakura:2014zwa, Sasakura:2014yoa}. 
However, there exist at least the following major differences from the statistical systems:
the signature of the exponent is opposite, and one can more freely take the integration domain, ${\cal C}_\phi$. 
Therefore, the dynamics would be much different. 
We leave this interesting aspect for future analysis.

From \eq{eq:simpwave} and \eq{eq:psifc}, it is obvious that
\[
\psi_{{\cal C}_\phi}(M)=\int_{{\cal C}_\phi} d\phi\ (\phi^2)^{\beta_N} (M\phi^3)^{-\gamma_N}
\label{eq:simpsol}
\]
is a solution. In this case, the dimension of $\phi$ is canceled in the integration, since $N+2 \beta_N-3\gamma_N=0$. 
Therefore, it should be the canonical form of the solution in the case that ${\cal C}_\phi$ is a finite domain. 

Let us compute the scaling dimension of the above solutions. We easily obtain 
\[
M_{abc}D^M_{abc} \psi(M)=-\gamma_N \psi(M)
\label{eq:scalepsiM}
\]
for all the solutions above, and this coincides with \eq{eq:ApplyDtoCkpm}.
Then the solutions in this section are dynamical except for the delicate case $N=1,2$ as discussed in Section \ref{sec:ransol}.

\section{Explicit examples}
\label{sec:explicitexample}

In this section, we will check the solutions for the simple cases with $N=1,2$.

\subsection{$N=1$}

For the $N=1$ case, taking $h(x)=1$ in \eq{eq:simpwave} gives
\[
\psi=\oint  \frac{d\phi}{\phi}=2 \pi i.
\]
Here the contour is assumed to be a closed path surrounding the origin.
This is actually the trivial solution to the constraint equation for $N=1$,
\[
D^M M D^M \psi=0.
\label{eq:wdwN=1}
\]

To obtain a non-trivial solution, we consider \eq{eq:simpwave} with $h(x)=\log (x)$:
\[
\psi_{\log}(M)=\oint\frac{d\phi}{\phi} \, \log \left(M\phi^3\right)=2 \pi i \log M + \hbox{const.},
\label{eq:simpwaveN=1log}
\]
which is indeed the other independent solution to \eq{eq:wdwN=1}.
Here the multi-valuedness of the logarithmic function in the 
integrand is {\it not} problematic: the first derivative, $h'(x)=\frac{1}{x}$, 
is a single-valued function, and one can safely justify the partial integration 
performed in \eq{eq:MDDsimpwave}.
Since $M D^M \psi_{\log} \neq 0$, this is a dynamical solution.

\subsection{$N=2$}

\begin{figure}[htbp]
\centering
\includegraphics[width=0.4\textwidth]{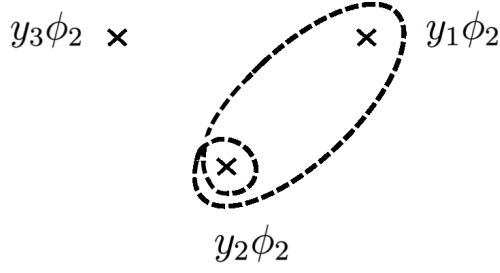}
\caption{A proper contour, $\mathcal{C}_{\phi_1}$: 
a closed dashed line winding two times around $\phi_1=y_2\phi_2$ 
and one time around $\phi_1=y_1\phi_2$.}
\label{cont}
\end{figure}

For the $N=2$ case, \eq{eq:simpsol} gives 
\[
\psi (M)=\int_{\mathcal{C}_{\phi}} d\phi_1 d\phi_2 
\frac{1}{\left( M_{abc} \phi_a\phi_b\phi_c \right)^\frac{2}{3}}.
\label{n2exsol}
\]
Let us check if one can really choose the contour, $\mathcal{C}_{\phi}$,
so that (\ref{n2exsol}) becomes a non-vanishing meaningful solution. 
To begin, when choosing the gauge,
\[
M_{111}=1, \ \ \ M_{112}=0, \ \ \ M_{122}=x_1, \ \ \ M_{222}=x_2,
\] 
one finds
\[
M_{abc}\phi_a\phi_b\phi_c&=\phi_1{}^3+3x_1\phi_1\phi_2{}^2+x_2\phi_2{}^3 \notag \\
&=(\phi_1-y_1\phi_2)(\phi_1-y_2\phi_2)(\phi_1-y_3\phi_2),
\]
where $y_i$'s are the three solutions to $y^3+3 x_1 y+x_2=0$.
We then implement the $\phi_1$-integration 
by choosing the contour, $\mathcal{C}_{\phi_1}$, in such a way as 
to enclose $\phi_1=y_2\phi_2$ two times and $\phi_1=y_1\phi_2$ one time (see Fig.\ref{cont}).
We take such a contour to make it consistent with 
the multi-valuedness of the integrand caused by the fractional power. 
This choice of contour is in fact democratic: 
every contour enclosing $\phi_1=y_i\phi_2$ ($i=1,2,3$) two times and 
$\phi_1=y_{j}\phi_2$ $(j\ne i)$ one time yield the same result up to a sign. 
After the $\phi_1$-integration, (\ref{n2exsol}) can be shown to be 
\[
\hbox{const.}\ \int_{\mathcal{C}_{\phi_2}} d\phi_2 \frac{1}{\phi_2 
\left[ (y_1-y_2)(y_2-y_3)(y_3-y_1) \right]^\frac{1}{3}}.
\label{afterphi1}
\]  
Choosing $\mathcal{C}_{\phi_2}$ to enclose $\phi_2=0$, one finally finds
\[
\psi (M) \propto \frac{1}{\left[ (y_1-y_2)(y_2-y_3)(y_3-y_1) \right]^\frac{1}{3}}\propto \frac{1}{(4 x_1^3+x_2^2)^\frac{1}{6}}\propto A(M)^{-\frac{1}{6}},
\]
where the last expression is valid in the whole space with 
$A(M)$ defined in \eq{eq:defAP}. This can be shown to be a kinematical solution, similarly to \eq{eq:psikin}.

\section{Solutions in $P$-representation without a cosmological constant}
\label{sec:solPrep}

The method in Section \ref{sec:simpsol} can also be used to solve 
the constraint equations in the $P$-representation. 
Let us start with an assumption,
\[
Q_{{\cal C}_\phi,h}(P)=\int_{{\cal C}_\phi} d\phi\ f(\phi^2) h(P\phi^3),
\label{eq:qfh}
\]
where $f$ will be determined shortly.

The momentum constraints can be shown to be satisfied by the same 
reasoning as in Section \ref{sec:simpsol}. 
For $\lambda=0$ in \eq{hH}, the Hamiltonian constraints are given by  
\[
(P_{abc} P_{bde} D^P_{cde}+\lambda_{H} P_{abb}) \psi(P)=0.
\label{eq:PPDpsi}
\]
As for the first term,
\[
P_{abc} P_{bde} D^P_{cde} Q_{{\cal C}_\phi,h}(P)
&=\int_{{\cal C}_\phi}  d\phi \ P_{abc}P_{bde}\phi_c\phi_d\phi_e f(\phi^2) h'(P\phi^3) \CR
&=\frac{1}{3} \int_{{\cal C}_\phi}  d\phi \ P_{abc} \phi_c f(\phi^2) D^\phi_b h(P\phi^3) \CR
&=-\frac{1}{3} \int_{{\cal C}_\phi}   d\phi\ D^\phi_b\left(P_{abc} \phi_c f(\phi^2)\right) h(P\phi^3) \CR
&=-\frac{1}{3}\int_{{\cal C}_\phi}  d\phi \ 
\left( P_{abb} f(\phi^2) +2 P_{abc}\phi_b \phi_c f'(\phi^2)\right) h(P\phi^3).
\label{eq:applyPPDQ}
\]
In the last line of \eq{eq:applyPPDQ},  the second term seems to be problematic 
for the constraint equations to be satisfied.
So, we set $f(x)=1$ to discard it. Then, from \eq{eq:applyPPDQ}, we obtain that 
\[
Q_{{\cal C}_\phi,h}(P)=\int_{{\cal C}_\phi} d\phi\ h(P\phi^3)
\label{eq:Qsol}
\]
satisfies
\[
\left( P_{abc} P_{bde} D^P_{cde} +\frac{1}{3} P_{abb} \right) Q_{{\cal C}_\phi,h}(P)=0.
\label{eq:Qwdw}
\]
The only difference of \eq{eq:PPDpsi} from \eq{eq:Qwdw} is the numerical coefficient of the second term,
and it is obvious that
\[
\psi_{{\cal C}_\phi,h}(P)=  Q_{{\cal C}_\phi,h}(P)^{3 \lambda_H}
\label{eq:psibyQ}
\]
satisfies the constraint equations \eq{eq:PPDpsi}. 

The scaling dimension can be computed as 
\[
P_{abc}D^P_{abc} Q_{{\cal C}_\phi,h}(P)
&=\int_{{\cal C}_\phi} d\phi\ \left(P \phi^3\right) h'(P\phi^3) \CR
&=\frac{1}{3} \int_{{\cal C}_\phi} d\phi\ \phi_a D^\phi_a h(P\phi^3) \CR
&=-\frac{N}{3} Q_{{\cal C}_\phi,h}(P).
\]
Thus,
\[
P_{abc} D^P_{abc} \psi_{{\cal C}_\phi,h}(P)
&=- N \lambda_{H} \psi_{{\cal C}_\phi,h}(P),
\label{eq:scalenlamda}
\]
which agrees with the necessary condition \eq{eq:criteriaP} for a kinematical solution.
The scaling dimension will become obvious, if we take $h(x)=x^{-\frac{N}{3}}$ as
\[
Q_{{\cal C}_\phi}(P)=\int_{{\cal C}_\phi} d\phi\ (P\phi^3)^{-\frac{N}{3}}.
\]
This should be a canonical expression when ${\cal C}_\phi$ is a finite domain,
since the dimension of $\phi$ is canceled in the integration.

Finally, let us check the type of the solution \eq{eq:psibyQ}.
By partial integrations, one obtains
\[
P_{acd}D^P_{bcd} Q_{{\cal C}_\phi,h}(P)&
=\int_{{\cal C}_\phi} d\phi \ P_{acd} \phi_b \phi_c \phi_d h'(P \phi^3) \CR
&=\frac{1}{3} \int_{{\cal C}_\phi} d\phi \ \phi_b D^\phi_a h(P\phi^3) \CR
&=-\frac{1}{3} \delta_{ab} Q_{{\cal C}_\phi,h}(P).
\]
This leads to $\hat {\cal J}_{(ab)}\psi_{{\cal C}_\phi,h} (P)=0$, namely, $\psi_{{\cal C}_\phi,h} (P)$ is a kinematical solution,
and might not be so interesting by itself.
However, it will be useful as a component to generate a variety of dynamical solutions in the following sense. 
Let us suppose that we have two solutions, 
say $\psi_1(P), \psi_2(P)$, which both satisfy the constraint equations \eq{eq:PPDpsi}. 
Then, it is obvious that the product,  $\psi_1(P)^\tau \psi_2(P)^{1-\tau}$, with an arbitrary number $\tau$ 
satisfies \eq{eq:PPDpsi}. Namely, one can construct one parameter family of solutions to the 
constraint equations from two independent solutions. This essentially comes 
from the fact that the Hamiltonian constraints are first-order differential 
operators in the $P$-representation. 
Now, a number of dynamical solutions in the $P$-representation can be obtained 
through the Fourier transform of the solutions in the $M$-representation in Section \ref{sec:ransol} and Section \ref{sec:simpsol}. 
Then, the above products of these solutions and $\psi_{{\cal C}_\phi,h} (P)$ will give families of dynamical solutions.

\section{Solutions with integration over matrix and tensor variables} 
\label{sec:solmatten}

The construction of the solutions in Section \ref{sec:solPrep} can be 
generalized to that with integration over matrix and tensor variables.
The first two solutions below are kinematical, while the final one is dynamical.

Let us consider
\[
Q_{{\cal C}_{K,\phi},h,g}(P)= \int_{{\cal C}_{K,\phi}} dK d\phi\ h(PK\phi) g(P\phi^3),
\]
where we have introduced an integration over a symmetric matrix 
$K$, in addition to the integration over $\phi$ as before,
$h,g$ are functions, and 
\[
PK\phi \equiv P_{abc} K_{ab}\phi_c.
\] 
By performing partial integrations as in the previous sections,
one can show that 
\[
P_{acd} & D^P_{bcd} Q_{{\cal C}_{K,\phi},h,g}(P) \CR
&=\int_{{\cal C}_{K,\phi}} dK d\phi\ P_{acd} \left\{\frac{1}{3} \left(2 K_{bc} \phi_d + K_{cd} \phi_b \right) h'(PK\phi) g(P\phi^3)
+\phi_b \phi_c \phi_d h(PK\phi) g'(P\phi^3) \right\} \CR
&=\frac{1}{3} \int _{{\cal C}_{K,\phi}}dK d\phi \left( 2 K_{bc} D^K_{ca} +\phi_b D^\phi_a \right) h(PK\phi)g(P\phi^3) \CR
&=-\frac{N+2}{3} \delta_{ab} \, Q_{{\cal C}_{K,\phi},h,g}(P),
\label{eq:linqckcp}
\]
where $D^K$ denotes partial derivatives with respect to $K$ as 
\[
D^K_{ab} K_{cd}=\frac{1}{2} \left( \delta_{ac}\delta_{bd} +\delta_{ad} \delta_{bc} \right).
\]
Then, 
\[
\psi_{{\cal C}_{K,\phi},h,g}(P)=Q_{{\cal C}_{K,\phi},h,g}(P)^{\frac{3 \lambda_H}{N+2}}
\]
satisfies the constraint equations \eq{eq:PPDpsi}.
\eq{eq:linqckcp} also shows that $\psi_{{\cal C}_{K,\phi},h,g}(P)$ 
is a kinematical solution.

A solution with integration over a tensor variable starts with considering
\[
Q_{{\cal C}_T,h}(P)=\int_{{\cal C}_T } dT\  h(PT),
\]
where $T$ is a symmetric  tensor with three indices, 
and
\[
PT \equiv  P_{abc}T_{abc}.
\]
By partial integrations as before, one can show that 
\[
P_{acd}  D^P_{bcd} Q_{{\cal C}_T,h} (P) =-\frac{(N+2)(N+1)}{6} \delta_{ab} Q_{{\cal C}_T,h}(P).
\label{eq:linpsit}
\]
Then
\[
\psi_{{\cal C}_T,h}(P)=Q_{{\cal C}_T,h}(P)^\frac{6 \lambda_H}{(N+1)(N+2)}
\]
can be shown to satisfy the constraint equations \eq{eq:PPDpsi}.
\eq{eq:linpsit} also shows that $\psi_{{\cal C}_T,h}(P)$ is a kinematical solution.

So far, in this and previous sections, we have only constructed 
kinematical solutions in the $P$-representation. 
A dynamical solution can be constructed by starting with the following integration over a symmetric matrix $K$,
\[
\label{KPPK_solnform}
Q_{{\cal C}_K,h}(P) = \int_{{\cal C}_K} dK\ h(KPPK),
\]
where 
\[
KPPK = K_{ab}P_{abc}P_{cde}K_{de}.
\]
As for the first term of the Hamiltonian constraints, 
\[
P_{abc}P_{bde}D^P_{cde} Q_{{\cal C}_K,h}(P)&= 2 \int_{{\cal C}_K} dK \ P_{abc}P_{bde} 
K_{\underline{c}\underline{d}}(PK)_{\underline{e}}h'(KPPK) \CR
&=\frac{2}{3} \int_{{\cal C}_K} 
dK \ (P_{abc}(PK)_b(PK)_{c}+2 P_{abc}P_{bde} K_{cd}(PK)_e)h'(KPPK) \CR
&=\frac{1}{3} \int_{{\cal C}_K} dK\ ((PK)_bD^K_{ab} +2 P_{abc}K_{cd}D^K_{bd})h(KPPK) \CR
&=-\frac{1}{3} \int_{{\cal C}_K} dK\ ((D^K_{ab}(PK)_b) +2 P_{abc}(D^K_{bd}K_{cd}))h(KPPK) \CR
&=-\frac{1}{3} \int_{{\cal C}_K} dK\ (P_{abb}+(N+1) P_{abb})h(KPPK) \CR
&=-\frac{N+2}{3} P_{abb} Q_{{\cal C}_K,h}(P),
\]
where the underlined indices are supposed to take the average over their permutations, 
and
\[
(PK)_a \equiv P_{abc}K_{bc}.
\]
In a similar way, the scaling dimension can be obtained as
\[
P_{abc}D^P_{abc}  Q_{{\cal C}_K,h}(P)=-\frac{N(N+1)}{2} Q_{{\cal C}_K,h}(P).
\]
These results show that 
\[
\psi_{{\cal C}_K,h}(P)=Q_{{\cal C}_K,h}(P)^\frac{3 \lambda_{H}}{N+2}
\label{eq:psickh}
\]
is a solution to the constraint equations, and its scaling dimension is given by 
\[
P_{abc}D^P_{abc} \psi_{{\cal C}_K,h}(P)=-\frac{3 N(N+1) \lambda_{H}}{2(N+2)} \, \psi_{{\cal C}_K,h}(P).
\label{eq:scalekh}
\]
By comparing with \eq{eq:criteriaP}, $\psi_{{\cal C}_K,h}(P)$ is a dynamical solution for $N>1$.

It would be meaningful to see whether $\psi_{{\cal C}_K,h}(P)$ is different from the solutions obtained in 
Section \ref{sec:ransol} and \ref{sec:simpsol}.
From \eq{eq:scalepsiM}, we obtain 
\[
-P_{abc}D^P_{abc} \psi(P)=i \hat P_{abc} \hat M_{abc} \psi=D^M_{abc} M_{abc}  \psi(M)
=\left( \frac{N(N+1)(N+2)}{6} -\gamma_N \right)  \psi(M)
\]
This differs from \eq{eq:scalekh} except for $N=1,4$. 
Therefore, $\psi_{{\cal C}_K,h}(P)$ is a new solution at least for $N\neq 1, 4$.

\section{Solutions in $P$-representation with a cosmological constant}
\label{sec:WDWcosmo}

The canonical tensor model with variables satisfying the generalized hermiticity 
condition does not allow a cosmological constant term,
because such a term violates the consistency of the first-class constraint algebra \cite{Sasakura:2012fb}.
On the other hand, if we restrict the variables to be real symmetric as in this 
paper, the algebra remains consistent even after such a cosmological term is introduced as in Section \ref{Introduction}. 
The Hamiltonian constraint equations with a cosmological constant $\lambda$ in the $P$-representation are given by
\[
(P_{abc}P_{bde} D^P_{cde}+\lambda_H^N P_{abb} - \lambda D^P_{abb})\psi(P)=0,
\label{eq:wdwcosmo}
\]
where the dependence of the normal ordering term on $N$ is explicitly written for later convenience.
The interpretation of $\lambda$ as a cosmological constant has been 
validated by comparing the classical dynamics of
the canonical tensor model with $N=1$ and the mini-superspace 
approximation of general relativity \cite{Sasakura:2014gia}.

In this section, we will explicitly construct a few series of solutions to 
\eq{eq:wdwcosmo} by generalizing the solutions obtained in the previous sections. 
Remarkably, it will be observed that all these solutions for $N=m$ can actually be
obtained from those to the constraint equations with {\it no} 
cosmological constant for $N=m+1$ by fixing the extra components of $P$.
In Section \ref{sec:generalcosmo}, we will prove a general theorem on this aspect.

Let us start with \eq{eq:applyPPDQ},
\[
P_{abc} P_{bde} D^P_{cde} Q_{{\cal C}_\phi,f,h}(P)
+\frac{1}{3} \int_{{\cal C}_\phi} d\phi \ \left[ P_{abb} f(\phi^2) 
+ 2 P_{abc}\phi_b \phi_c f'(\phi^2)\right] h(P\phi^3)=0,
\label{eq:startWDW}
\]
where $Q_{{\cal C}_\phi,f,h}(P)$ is assumed to have the 
expression on the righthand side of \eq{eq:qfh}. By performing a 
partial integration, the last term can be computed to be
\[
\frac{2}{3} \int_{{\cal C}_\phi} d\phi \  P_{abc}\phi_b \phi_c f'(\phi^2) h(P\phi^3)
&=\frac{2}{9}\int_{{\cal C}_\phi} d\phi \  f'(\phi^2) D^\phi_a h^I(P\phi^3) \CR
&=-\frac{4}{9}\int_{{\cal C}_\phi} d\phi \  \phi_a f''(\phi^2) h^I(P\phi^3) , 
\label{eq:compthird}
\]
where $h^I(x)$ is a function satisfying  
\[
\frac{d}{dx} h^I(x)= h(x).
\]
Now let us assume
\[
&h^I(x)=\frac{1}{A}\exp[A x],
\label{eq:hisexp} \\
&f''(x)=B x f(x),
\label{eq:condfx}
\]
with numerical constants, $A,B$. 
Then \eq{eq:compthird} can further be computed as
\[
-\frac{4}{9}\int_{{\cal C}_\phi}d\phi \  \phi_a f''(\phi^2) h^I(P\phi^3) 
&=-\frac{4 B }{9 A} \int_{{\cal C}_\phi}d\phi \ \phi_a \phi^2 f(\phi^2) h(P \phi^3) \CR
&=-\frac{4B}{9A^2} D^P_{abb} \int_{{\cal C}_\phi} d\phi \ f(\phi^2) h(P \phi^3).
\]
This concludes
\[
\left[ P_{abc} P_{bde} D^P_{cde} +\frac{1}{3} P_{abb}
-\frac{4B}{9A^2} D^P_{abb}\right] Q_{{\cal C}_\phi,f,h}(P)=0.
\label{eq:WDWwithcos}
\]
Thus, if we consider
\[
\psi_{{\cal C}_\phi,f,h}(P)= Q_{{\cal C}_\phi,f,h}(P)^{3 \lambda_{H}^N},
\label{eq:relpsiQP}
\]
we obtain
\[
\left[P_{abc} P_{bde} D^P_{cde}+\lambda_{H}^NP_{abb} 
-\frac{4B}{9A^2} D^P_{abb}\right]\psi_{{\cal C}_\phi,f,h}(P)=0.
\]
Namely, $\psi_{{\cal C}_\phi,f,h}(P)$ satisfies the constraint equations with a cosmological constant,
\[
\lambda=\frac{4B}{9A^2}.
\]

In fact, the solution to \eq{eq:condfx} is given by the Airy function,
\[
f(x)=\hbox{Airy}\left[ B^\frac{1}{3} x \right]
=\int_{{\cal C}_z} dz\ \exp \left[ -B^\frac{1}{3} x z +\frac{z^3}{3} \right].
\label{eq:fairy}
\]
Then, by putting $h(x)=\exp (Ax)$ and \eq{eq:fairy} into the expression \eq{eq:qfh}, 
one obtains an intriguing expression,
\[
Q_{{\cal C}_\phi,f,h}(P)&=\int_{{\cal C}_{\phi,z}}  d\phi dz \ 
\exp \left[A\, (P\phi^3)-B^\frac{1}{3} \phi^2 z +\frac{z^3}{3} \right] \CR
&=\hbox{const.} \int _{{\cal C}_{\phi,z}}d\phi dz \ \exp \left[P\phi^3-\phi^2 z +\frac{A^2}{3 B} z^3 \right] \CR
&=\hbox{const.} \int _{{\cal C}_{\phi}} \prod_{a=1}^{N+1} d\phi_a \exp\left[\tilde P\phi^3 \right]
\]
with
\[
&\tilde P_{abc}=P_{abc}, \CR
&\tilde P_{ab\,N+1}=-\frac{1}{3}\delta_{ab}, 
\label{eq:lambdaN+1}\\
&\tilde P_{a\,N+1\, N+1}=0, \CR
&\tilde P_{N+1\, N+1\,N+1}=\frac{4}{27 \lambda},
\nonumber
\]
where we have rescaled $\phi,z$, and have renamed $z=\phi_{N+1}$.
Thus, $Q_{{\cal C}_\phi,f,h}(\tilde P)$ has the form as \eq{eq:Qsol}, 
and therefore $\psi_{{\cal C}_\phi,f,h}(\tilde P)=Q_{{\cal C}_\phi,f,h}(\tilde P)^{3 \lambda_{H}^{N+1}}$ actually 
satisfies the constraint equations with no cosmological constant for the variables $\tilde P$, as 
same as the wave-function in \eq{eq:psibyQ}.
This shows that the cosmological constant can be absorbed 
into some of the dynamical variables by increasing $N$.

Below, we will show that a similar fact holds also for a wider class of solutions. 
Let us consider the solution, $Q_{{\cal C}_\phi,h}(P)$ in \eq{eq:Qsol}, 
and start with \eq{eq:Qwdw},
\[
P_{abc}P_{bde} D^P_{cde} Q_{{\cal C}_\phi,h} (P)
+\frac{1}{3} P_{abb} Q_{{\cal C}_\phi,h}(P)=0,
\label{eq:startWDWQP}
\]
where $a,b,c=1,2,\cdots, N+1$.
Let us divide the index set as 
\[
a=\left\{ 
\begin{array}{l}
i\in I_{N}=\{1,2,\ldots,N\}\\
z=N+1
\end{array}
\right.
,
\label{eq:indexdivision}
\]
and consider a subspace in which $P$'s containing the index $z$ are fixed as
\[
&P_{zzz}=A, \CR
&P_{zzi}=0,
 \label{eq:assumptionP}\\
&P_{zij}=B \delta_{ij},\nonumber
\]
where $i,j\in I_{N}$. 
Below, $i,j,\ldots$ will be used for the elements in 
$I_N$, and $a,b,\ldots$ for both $I_N$ and $z$.

On the subspace \eq{eq:assumptionP},  the second term of \eq{eq:startWDWQP} for $a=i\in I_N$ becomes 
\[
\frac{1}{3} P_{ibb} Q_{{\cal C}_\phi,h} (P)=\frac{1}{3} P_{ijj} Q_{{\cal C}_\phi,h} (P).
\label{eq:comppibb}
\]
On the subspace, the first term of \eq{eq:startWDWQP} for $a=i$ can be evaluated as  
\[
&P_{ibc}P_{bde}D^P_{cde} Q_{{\cal C}_\phi,h} (P)
=\int_{{\cal C}_\phi} d\phi \ P_{ibc}P_{bde} \phi_c \phi_d \phi_e h'(P\phi^3) \CR
\ \ \ \ &=
\int_{{\cal C}_\phi}  d\phi\ [
P_{ijk}P_{jlm} \phi_k \phi_l \phi_m
+3 B P_{ijk} \phi_j \phi_k  \phi_z 
+(2 B^2+AB)  \phi_i (\phi_z)^2 +B^2  \phi_i \phi_j^2
]
h'(P\phi^3),
\label{eq:PPDQphi}
\]
where it is important to notice that \eq{eq:assumptionP} do not contain any conditions on the corresponding derivatives, $D^P_{zab}$.
The following two identities hold for \eq{eq:assumptionP}:
\[
&0=\int_{{\cal C}_\phi} d\phi\ D^\phi_i \left( \phi_z h(P\phi^3)\right) 
=\int _{{\cal C}_\phi}d\phi\ \left[ 3 P_{ijk} \phi_j \phi_k \phi_z 
+ 6 B \phi_i \phi_z^2 \right] h'(P \phi^3) ,\CR 
&0=\int_{{\cal C}_\phi} d\phi\ D^\phi_z\left( \phi_i h(P\phi^3)\right)
= \int_{{\cal C}_\phi} d\phi \ \left[ 3 A \phi_i (\phi_z)^2 
+ 3 B \phi_i \phi_j^2 \right] h'(P\phi^3).
\]
These identities can be used to delete the terms, 
$P_{ijk} \phi_j \phi_k\phi_z $ and $\phi_i (\phi_z)^2$ in \eq{eq:PPDQphi}, to obtain
\[
P_{ibc}P_{bde}D^P_{cde} Q_{{\cal C}_\phi,h} (P)&=\int_{{\cal C}_\phi} d\phi\ \left[
P_{ijk}P_{jlm} \phi_k \phi_l \phi_m
+\frac{4 B^3}{A} \phi_i \phi_j^2  \right] h'(P\phi^3) \CR
&=\left[
P_{ijk}P_{jlm} D^P_{klm}
+\frac{4 B^3}{A} D^P_{ijj} \right] Q_{{\cal C}_\phi,h} (P)
\]
Thus the wave-function \eq{eq:psibyQ} satisfies the constraint equations with a cosmological constant,
\[
\lambda=-\frac{4 B^3}{A}.
\]

\section{A theorem for ignoring a cosmological constant}
\label{sec:generalcosmo} 

In this section, as generalization of Section \ref{sec:WDWcosmo},
we will prove that a solution to the constraint equations with 
no cosmological constant for $N=m+1$ can always generate a solution to the 
constraint equations with a cosmological constant for $N=m$
by fixing the extra components of $P$ as in \eq{eq:assumptionP} with $A=B$. 
 
Let us divide the index set as in \eq{eq:indexdivision}, and assume 
that $i,j,\dots$ denote the elements in $I_N$, while $a,b,\ldots$ both $I_N$ and $z$.
Let us assume that a wave-function $\psi(P)$ satisfies the 
constraint equations with no cosmological constant 
and the momentum constraints as 
\[
&\left[ P_{abc}P_{bde} D^P_{cde} +\lambda_{ H}^{N+1} P_{abb} \right] \psi(P)=0, 
\label{eq:WDWpsi}\\
&\left[ P_{acd} D^P_{bcd} - P_{bcd} D^P_{acd} \right] \psi(P)=0,
\label{eq:oninv}
\]
where the dependence of the normal ordering term on $N+1$ is explicitly written.
Let us assume \eq{eq:assumptionP} for $P_{zab}$, while $P_{ijk}$ are left arbitrary.  
Then, on the subspace, the second term in \eq{eq:WDWpsi} for $a=i$ is given by
\[
P_{ibb}=P_{ijj}.
\label{eq:piaa}
\]
As for the first term in \eq{eq:WDWpsi}, by putting \eq{eq:assumptionP}, we obtain
\[
P_{ibc} P_{bde}D^P_{cde} \psi(P)&=\left[ P_{ijk} P_{jde}D^P_{kde}
+ P_{ijz} P_{jde}D^P_{zde}+P_{izj} P_{zde}D^P_{jde} \right] \psi(P) \CR
&=\left[ P_{ijk} P_{jlm} D^P_{klm}+2 P_{ijk} P_{jlz} D^P_{klz}
+P_{ijz} P_{jkl}D^P_{zkl}+2 P_{ijz} P_{jkz}D^P_{zkz} \right. \CR
&\ \ \ \ \ \ \ \ \left. + P_{izj} P_{zkl}D^P_{jkl}+P_{izj} P_{zzz}D^P_{jzz} \right] \psi(P) \CR
&=\left[ P_{ijk} P_{jlm} D^P_{klm}+3 B P_{ijk} D^P_{jkz}+(2 B ^2+AB) D^P_{izz}+ B^2  D^P_{ijj} \right] \psi(P).
\label{eq:ppdpqpgen}
\]
On the other hand, by putting \eq{eq:assumptionP} into \eq{eq:oninv} for $a=i,b=z$, we obtain
\[
0&=\left[ P_{iab} D^P_{zab} - P_{zab} D^P_{iab} \right] \psi(P)  \CR
&=\left[ P_{ijk} D^P_{zjk}+2 P_{ijz} D^P_{zjz} - P_{zjk} D^P_{ijk}-P_{zzz} D^P_{izz} \right] \psi(P) \CR
&=\left[ P_{ijk} D^P_{jkz}+(2 B-A) D^P_{izz} - B D^P_{ijj}\right] \psi(P).
\label{eq:momrel}
\]
By adding \eq{eq:momrel} multiplied by a free parameter $t$,  one finds that
the last line of \eq{eq:ppdpqpgen} is equivalent to 
\[
[ P_{ijk} P_{jlm} D^P_{klm}+(3 B+t)  P_{ijk} D^P_{jkz}+\{ 2 B ^2+AB +t (2B-A) \} D^P_{izz}+(B^2- t B)  D^P_{ijj} ] \psi(P).
\]
Here, by choosing $t=-3B,\ A=B$, 
one can delete the second and third terms to obtain 
\[
P_{ibc} P_{bde}D^P_{cde} \psi(P)=\left[ P_{ijk} P_{jlm} D^P_{klm}+ 4 A^2  D^P_{ijj} \right] \psi(P).
\label{eq:ppdeval}
\]
Then, from \eq{eq:WDWpsi}, \eq{eq:piaa} and \eq{eq:ppdeval}, we conclude 
\[
\left[
P_{ijk} P_{jlm} D^P_{klm}+ 4 A^2  D^P_{ijj} +\lambda_{H}^{N+1} P_{ijj}
\right] \psi(P)=0
\]
holds on the subspace \eq{eq:assumptionP}. 
Finally, we define
\[
\psi_N(P)=\psi(P)^{\lambda_H^N/\lambda_H^{N+1}}.
\]
Then, $\psi_N(P)$ satisfies the constraint equations,
\[
\left[
P_{ijk} P_{jlm} D^P_{klm}+ 4 A^2  D^P_{ijj} +\lambda_{H}^{N} P_{ijj}
\right] \psi_N(P)=0,
\]
with a cosmological constant,
\[
\lambda=-4 A^2.
\]

Two comments are in order. One is that, to obtain a positive 
cosmological constant, we have to perform analytic continuation 
of $P_{zzz}=P_{zii}=A$ to pure imaginary values.
On the other hand, the explicit general solutions with a cosmological constant for $N=2,3$ in Section \ref{n2} and \ref{n3}
show that $\lambda$ appears without such imaginary numbers. 
This suggests that there might be another more improved discussion to ignore a cosmological constant than above.
The second is that it is totally unclear 
whether all the solutions to the constraint equations with a 
cosmological constant can be obtained from those with no cosmological constant by the 
present method of increasing $N$. What we have shown is merely the reverse; the 
latter to the former. Therefore, at present, we do not know whether 
we can totally ignore a cosmological constant in the analysis of the 
canonical tensor model. 
We leave this interesting possibility for future study.

\section{Summary and discussions}
\label{discussions}

In this paper, we studied exact physical states in the canonical tensor model. 
Our interest in this paper was to study the canonical tensor model of higher $N$ 
with and without a cosmological constant, and wherever possible to find the 
exact general solutions to the constraint equations. 
After introduction, we presented an outline of the canonical tensor model 
in Section \ref{Introduction}. Here we described the 
constraint algebra and the corresponding equations satisfied by the physical states
of the canonical tensor model. Here we also introduced the concept of
kinematical and dynamical parts of the states,
where the latter reflects the non-linear characteristics of the constraints and would be physically more relevant.
We then 
proceeded to study the canonical tensor model of various $N$. 
The achievements of this paper can be summarized into the following three categories.

The first is that we have explicitly obtained the physical wave-functions for $N=2,3$, and have studied their features.
We considered the cases with and without a cosmological 
constant, and explicitly wrote down and solved the set of the partial differential equations representing the constraints. 
While the cases without a cosmological constant were more or less straightforward to solve in a standard method
 (was indeed previously solved for $N=2$ in \cite{Sasakura:2013wza}),
the cases with a cosmological constant set a new challenge.
In order to solve them, we solved on an $S_2\, (S_3)$-symmetric subspace 
in the configuration space for $N=2\, (N=3,\hbox{resp.})$ and then extended the solution over the whole space
for $N=2$.
Remarkably, the solutions in both the cases with and without a cosmological constant 
turned out to have the same form when expressed with some functions 
in which all the effects of the cosmological constant are included.
This actually hints that the inclusion of the cosmological constant does not change the structure 
of the system and tickles our intuition that this can be true even for higher-$N$ tensor models. 
In order to investigate this issue we went through a different path in the last sections.

Since we did not assume any boundary (or initial) conditions, the solutions contain undetermined functions.
However, we were able to study the possible locations of the peaks of the wave-functions,
which would be physically interpreted as enhanced configurations. 
We have found that, for $N=3$ with no cosmological constant, 
the configurations invariant under an $SO(2)$ symmetry as well as those associated to a Grassmann algebra 
(this also appears for $N=2$) can be enhanced. 
Moreover, the intersecting points of these two cases have additionally a Lorentz boost symmetry. 
While the $SO(2)$ and Lorentz boost transformations commute with each other for $N=3$, and thereby do not give rise to 
any interesting pictures, we expect more non-trivial phenomena to arise in higher-$N$ tensor models. 

The second is that we have systematically constructed some series of solutions valid for any $N$ from the perspective of 
statistical systems on random networks, or random tensor networks.
It was noticed that the canonical tensor model shares some resemblances
with the statistical systems on random networks, where the 
Hamiltonian vector flow of the canonical tensor model corresponds to the 
renormalisation group flow of the statistical system \cite{Sasakura:2014zwa}. 
The ``grand-type'' partition function of the statistical system \eq{eq:zkmc} satisfies the momentum constraints
of the tensor model. However, it doesn't satisfy the Hamiltonian 
constraints. 
This was cured by integrating the ``grand-type'' partition function in a scale-free manner,
which provided a physical wave-function of the canonical tensor model with no cosmological constant.
This is a remarkable achievement, as it gives a series of solutions valid for any $N$. 
An important fact noted in the analysis is that 
the solutions constructed in this manner are dynamical at least for $N\ge3$. 
An interesting connection between the ``grand-type'' partition function and an extension 
of Airy function was also noticed by transforming the integration variables.  

The above process of constructing physical wave-functions was simplified and generalized. 
We have constructed more general solutions valid for any $N$ and no cosmological constant
in terms of simpler expressions with integration over an $N$-component vector.
These solutions are dynamical at least for $N \geq 3$ as above. 
A similar process of constructing wave-functions for no cosmological constant 
has been performed in the conjugate (momentum) representation. 
However, these solutions turned out to be kinematical.
Then we also considered integration over matrix and tensor variables,
and found a new dynamical solution as well as kinematical solutions.
We pointed out that,
though kinematical solutions might be of less physical interest than dynamical ones,
the former can extend the variations of the latter with continuous parameters.
 
The third is that we have performed a general analysis of the wave-functions in the momentum representation
for the cases with a cosmological constant.
We started with the similar assumptions about the forms of wave-functions and solved the constraint equations.
Remarkably, the solutions for $N=m$ with a cosmological constant turned out to be obtained by restricting 
the domains of the solutions for $N=m+1$ with no cosmological constant.
Then we have proven a general theorem that the latter can always generate the former by such restriction.
This suggests that the effect of the cosmological constant can be absorbed into the dynamics of the tensor model with increased $N$.
This is somewhat like in general relativity, where a de Sitter 
background (maybe arising due to a cosmological constant) 
can be studied by embedding it in a flat space-time of
one higher dimension. This property is also true for 
more general solutions. 

It is now evident that there exist at least some tractable solutions valid for any $N$ in the canonical tensor model,
though the quantized constraints on physical states are non-linear and complicated.
It would be an interesting question how further we can proceed: is it possible to write down 
all the solutions in some tractable manners? On the other hand, from the physical point of view, it would rather be important to obtain
a unique physical state which satisfies a physically required initial condition. 
Then, one can discuss not only the possibilities but the real configurations
which become the peaks of the wave-function. This will inevitably be related with
the space-time emergence in the canonical tensor model, and as explicitly shown for $N=3$, there would be
symmetry enhancement, which would be related with why the universe is a homogenous space-time with various gauge symmetries.
This enhancement would also suggest possible connections between the canonical tensor model and the group field theories\cite{Carrozza:2014rba,Rivasseau:2013uca,Carrozza:2013mna,
Geloun:2013zka,Geloun:2013saa,Samary:2013xla,Carrozza:2013wda,
Carrozza:2012uv,Geloun:2012bz,BenGeloun:2012pu,BenGeloun:2011rc,Boulatov:1992vp,Ooguri:1992eb,
DePietri:1999bx,Freidel:2005qe,Oriti:2011jm}, in which 
Lie groups are embedded as input structures.  
Then, what is the initial condition? This question would also be tightly related to what is time in the canonical tensor model,
which contains no preferred time variable in its basic formulation. 
Finally, the discussions on a cosmological constant in this paper 
strongly suggests that it can totally be ignored.
This would be a really interesting possibility in view of the so-called cosmological constant problem,
and we hope to complete the discussions in future study.

\bigskip
\centerline{\bf Acknowledgements} 

GN would like to thanks NS, Shinya Aoki and others at 
the Yukawa Institute for Theoretical Physics (YITP) in Kyoto University
for providing the wonderful hospitality where a part of the work was done,
and is grateful to them for providing the generous financial support.
YS is very grateful to YITP, where a part of this work was done, for their warm hospitality, 
and would like to acknowledge financial support from YITP: 
the bilateral international exchange program (BIEP), 
and the long-term international exchange program for young researchers.

\end{document}